\documentclass[useAMS,usenatbib]{mn2e}

\usepackage{graphicx}
\usepackage{scalefnt}
\usepackage{amssymb}
\usepackage{upgreek}
\usepackage{amsmath}
\usepackage{balance}
\usepackage{textcomp}
\usepackage{ulem}
\usepackage{float} 

\def \spose #1{\hbox to 0pt{#1\hss}}
\def \lta {\mathrel{\spose{\lower 3pt\hbox{$\mathchar"218$}}
     \raise 2.0pt\hbox{$\mathchar"13C$}}}
\def \gta {\mathrel{\spose{\lower 3pt\hbox{$\mathchar"218$}}
     \raise 2.0pt\hbox{$\mathchar"13E$}}}
\def \kms {{km s$^{-1}$}}

\def \gastwo {\textsc{ESF-gasoline2}}

\def \msun {\ifmmode \rm M_{\odot}\else $\rm M_{\odot}$ \fi}
\def \rvir {\ifmmode \rm R_{\rm vir}\else $\rm R_{\rm vir}$\fi}
\def \mvir {\ifmmode \rm M_{\rm vir}\else $\rm M_{\rm vir}$\fi}
\def \Mvir {\ifmmode \rm M_{\rm vir}\else $\rm M_{\rm vir}$\fi}
\def \ovi {{\sc Ovi}}
\def \hi {{\sc Hi}}

\newcommand{\ergsec}	{\ifmmode \mathrm{\,erg~s}^{-1} \else \,erg~s$^{-1}$\fi}
\newcommand{\yr}	    {\ifmmode \mathrm{yr} \else yr\fi}
\newcommand{\mpc}	{\ifmmode \,\mathrm{Mpc}^{-3} \else \,Mpc$^{-3}$\fi}
\newcommand{\Msun}	{\ifmmode \,\mathrm M_{\odot} \else $\,\mathrm M_{\odot}$\fi}
\newcommand{\Mhalo}	{\ifmmode M_{\mathrm{Halo}} \else $M_{\mathrm{Halo}}$\fi}
\newcommand{\Mstar}	{\ifmmode {M}_\star \else ${M}_\star$\fi}

\title[NIHAO CGM]{NIHAO VIII: Circum-galactic medium and outflows -
The puzzles of \hi\, and \ovi\, gas distributions}
\author[T. A. Gutcke et al.]{Thales A. Gutcke$^{1}$\thanks{thales@mpia.de}, Greg S. Stinson$^{1}$,
  Andrea V. Macci\`o$^{2,1}$, Liang Wang$^{3}$,
\newauthor{Aaron A. Dutton$^{2}$} \\
$^1$Max-Planck-Institut f\"ur Astronomie, K\"onigstuhl 17, 69117
Heidelberg, Germany\\
$^2$New York University Abu Dhabi, PO Box 129188, Abu Dhabi, UAE\\
$^3$Purple Mountain Observatory, the Partner Group of MPI f\"ur Astronomie, 2 West Beijing Road, Nanjing 210008, China\\}

\begin{document}

\pagerange{\pageref{firstpage}--\pageref{lastpage}} \pubyear{---}

\maketitle

\label{firstpage}

\begin{abstract}


We study the hot and cold circum-galactic medium (CGM) of 86 galaxies of the
cosmological, hydrodynamical simulation suite NIHAO.  
NIHAO allows a study of how the $z=0$ CGM varies across 5 orders of magnitude of stellar mass
using \ovi\, and \hi\, as proxies for hot and cold gas. 
The cool \hi\, covering fraction and column density profiles match
observations well, particularly in the inner CGM. 
\ovi\, shows increasing column densities with mass, a trend seemingly echoed in the observations.
As in multiple previous simulations, 
the \ovi\, column densities in simulations are lower than observed
and optically thick \hi\, does not extend as far out as in
observations. We take a look at the collisional ionisation fraction of
\ovi\, as a function of halo mass. 
We make observable predictions of the bipolarity of outflows and their
effect on the general shape of the CGM. 
Bipolar outflows can be seen out to around 40 kpc {in intermediate and
low mass halos ($\Mhalo<10^{11}\Msun$), but outside that radius,
the CGM is too well mixed to detect an elongated shape. Larger halos
have extended gas discs beyond the stellar disc that dominate the
shape of the inner CGM.}
The simulated CGM is remarkably spherical even in low mass simulations.
The chemical enrichment of both halo and disc gas follow expected increasing trends 
as a function of halo mass that are well fit with power laws. 
These relations can be used in non-hydrodynamic models, such as
semi-analytic models.

\end{abstract}


\section{Introduction}

The gas surrounding galaxies, the ``circum-galactic medium'' (CGM), 
plays a major role in galaxy formation and  evolution.  Gas flows 
into the bottom of potential energy wells and turns into stars to create galaxies.
As gas falls into galaxies, the mass of the dark matter halo determines its velocity
and the gas heats to the virial temperature.
\citet{Rees1977} showed that the cooling time of the CGM gas determines
how much gas cools onto the galaxy, which corresponds 
to the mass scales at which galaxies form.

Theorists are still working out the details of the geometry of that collapse.
\citet{White1978} {included dark matter in the
  collapse and introduced the idea of hierarchical
  merging to form the most massive halos}.
\citet{Keres2005} pointed out that when galaxy masses are low, gas  
first cools onto filaments before flowing into galaxies along those narrow paths.
\citet{Birnboim2003} showed that as the galaxy increased in mass,
those filaments would shock near the virial radius and the CGM would 
cut off the cold gas supply to the galaxy.  The shocking of accreting gas
{is one of the many attempts at an explantation of } why star formation becomes so inefficient in galaxies more
massive than the Milky Way.

In such models, a large fraction of the baryons in galaxies around and above
$10^{11.5} \msun$ {in halo mass} is distributed in the CGM.  However, since it is
spread out over such a large volume it is diffuse enough that 
it does not emit much radiation.
What radiation is emitted comes out at hard to observe energies like
X-rays \citep[see e.g.][]{White1991, Crain2010}.  It is thus incumbent
on modelers to make predictions about the chemical phase, orientation
and shapes of the CGM in their models. 

Fortunately, observers can detect diffuse gas in absorption if a bright source like a quasar
lies behind it.  
Many absorption lines useful for diagnostics are in the UV portion of the spectrum.  
Thus, it is easiest to study the CGM at high redshifts, 
such that the lines redshift into the optical portion of the spectrum
\citep[e.g.][]{Cowie1995, Schaye2003, Hennawi2006,Steidel2010,Rudie2012, Lehner2014}.  

UV spectrographs were added to the {\it Hubble Space Telescope} (HST),
enabling studies of the low redshift CGM \citep[e.g.][]{Thom2008,Tripp2008,Prochaska2011}.
The {\it Cosmic Origins Spectrograph} (COS) added far-UV sensitivity and
made studies of the low redshift CGM easier.

The COS-Halos project focused on studying the CGM surrounding a large sample of 
$\sim L^\star$ galaxies \citep{Tumlinson2013}.  
\citet{Danforth2014} also studied the low redshift inter-galactic
medium (IGM).
COS-Halos studied {\sc Ovi} as one of its highly ionized species.  
Oxygen {is collisionally ionised to} {\sc Ovi} in gas between  $10^5 < T/K < 10^6$.  
{It is also possible for lower temperature gas to be photoionised
  to  {\sc Ovi} by an external radiation field.}
Oxygen is produced primarily in type II supernovae (SNII) and can flow out
of the galaxy with the galactic wind.
Thus, {\sc Ovi} can be either a tracer of warm/hot gas in the CGM or a cooler
phase that may have flowed out in the wind.

{\citet{Tumlinson2011} found {\sc Ovi} detections (their column
density detection limit being $N_{\rm OVI} \gtrsim 10^{14} {\rm cm}^{-2}$) in  the majority of their 30 observed star forming galaxies, 
but in none of their sample of red galaxies. }
In galaxies with {\sc Ovi} detections, it was extended in projection out to 300 kpc, 
beyond even the virial radius.
Making typical assumptions about the metallicity (0.5 $Z_\odot$),
abundance ratios (solar), and ionization field shining on the CGM
leads to high oxygen masses in the CGM, comparable to the amount
of oxygen found in the Milky Way disc \citep{Tumlinson2011, Peeples2014}.

Detailed examination of the absorption line profiles showed hints of components
with narrow lines embedded inside broader overall profiles \citep{Tripp2008,Thom2008}.  
{These tentative findings hint towards {\sc Ovi} being cooler than
expected by collisional ionisation models and generally more
consistent with being photoionised.}

One surprising result of COS-Halos was  that a large fraction of the CGM mass
is cool and in a low ionization state.  The gas also appears to be in dense knots
that are not in hydrostatic equilibrium with their surroundings \citep{Werk2014}.
How such gas is supported in the CGM remains a mystery.

Local L$^\star$ galaxies are not the only place a surprising amount of cool gas has been found 
\citep[for other places, see][]{Rudie2012,Martin2012, Prochaska2013,
  Crighton2013}.

Several studies have looked at the {\sc Hi} covering fraction at a variety of masses throughout
cosmic time \citep{Rudie2013, Prochaska2014}.  
Commonly, they report the fraction of sight lines in which {\sc Hi} is detected.
\citet{Fumagalli2014} provides a nice summary.
While models generally predict that the CGM of {luminous
  galaxies (in modelling generally assumed to be the more massive galaxies) } is comprised of more hot gas,
the observations find higher {\sc Hi} covering fractions of the CGM
surrounding {more luminous} galaxies.

{Some work has also been focused on understanding the geometrical
  extent of the CGM.} \citet{Bordoloi2014} present an investigation of the geometry of
cool outflows traced by MgII in the absorption line spectra of 486 zCOSMOS
galaxies. Splitting their sample by inclination, they
find a higher equivalent width out to 40-50 kpc for galaxies with near face on
inclination than for edge on galaxies. They take this to be indicative
of bipolar outflows perpendicular to the plane of the disc, since
absorption spectra will capture the outflowing component face on more
strongly than edge on. {\citet{Kacprzak2015}, using \ovi\, absorption lines, present a bimodality in the azimuthal angle distribution
  around 53 HST-imaged galaxies. They conclude that this bimodality is
  consistent with minor-axis driven outflows and that \ovi\, is not
  mixed throughout the CGM.}

\subsection{Numerical predictions and observations}

Galaxy formation models have used star formation feedback in an
attempt to explain the low efficiency of star formation in galaxies
\citep{Springel2003, Murray2005, Oppenheimer2006, DallaVecchia2008, 
Dave2011, Hopkins2012, Puchwein2013, Vogelsberger2013, Kannan2014}. 
Stellar-driven winds could not only deplete the central galaxy of gas, its fuel for star formation,
but also transport metal-enriched gas to large distances from the
stars \citep[e.g.][]{Aguirre2001, Theuns2002, Dave2007, Cen2011,Christensen2015}. 

Numerical simulations commonly use one of two prescriptions to model outflows: 
{kinetic} \citep{Springel2003, Dave2007, Oppenheimer2010, Ford2015, Suresh2015} 
and thermal \citep{Stinson2006, DallaVecchia2012, Shen2012, Hummels2013, Schaye2015, Rahmati2015}.  
The {kinetic} model uses a prescription for wind velocities launched from discs.
Thermal models make gas around star formation regions hot and allow the supersonic
pressure of the hot gas to push gas out of galaxies.
\citet{Durier2012} find that {kinetic wind} and thermal models converge to the same behavior using correct hydrodynamics.

{Kinetic wind models} typically turn off hydrodynamics of the wind particles including 
the recent {\it Illustris} simulations \citet{Vogelsberger2014}.  (But see \citet{Schaye2010} for an implementation of a wind model
without hydrodynamical decoupling.)
Early studies suggested that without hydrodynamics, 
such wind prescriptions preserved convergence in results of different resolutions.
It is unclear whether such convergence remains in the most recent studies \citep{Vogelsberger2013}.

Many simulated galaxies now broadly match the low redshift stellar mass-halo mass relation
\citep{Oppenheimer2012,Aumer2013,Stinson2013,Hopkins2014,Vogelsberger2014, Christensen2015,Schaye2015,Wang2015}. 
It is thus worth making a comparison with observations that could lend insight into how the CGM formed 
and whether it is the repository for the missing baryons in the Universe.

\subsubsection{{\sc Ovi}}

{Hydrodynamical simulations that compare their \ovi\,
  \citep[e.g.][]{Hummels2013, Ford2015, Suresh2015} with
  COS halos observations \citep{Tumlinson2011} find that their
  fiducial models produce nearly a dex lower column densities across
  impact parameters and, when compared, also across luminosities,
  than the observations.  }

 {Our earlier look using a thermal model \citep[][henceforth S12]{Stinson2012},
showed good agreement with \ovi\, observations, although this was a
small sample of two galaxies. The purpose of this work is to expand
that sample.}

Using an Eulerian grid code, \citet{Hummels2013} studied variations of
the CGM with thermal stellar feedback
strength. While they were able to match many ion species column densities, they found that \ovi\,
presented the biggest challenge.  
Various strengths of their thermal feedback scheme changed the typical
\ovi\, column densities by a half an order of magnitude and the extent of the \ovi\,
halo by an order of magnitude. {These changes can be attributed to
  changes in star formation rates due to the availability of gas
  reservoirs, but also to the differences in outflow strength which
  effect the metallicity distribution.}
The feedback prescription in which cooling was suppressed produced results 
most consistent with observations.
Even in that simulation, the simulated \ovi\, column densities remain about 
half a dex too low.  
In terms of radial extent, the cooling suppressed feedback also extends as far as the observations.
Plenty of oxygen was present in their gaseous halos,
but the temperature range for \ovi\, is small enough that not much gas
stayed in that state, assuming collisional ionisation dominates.

{Focusing on the effect of feedback, \citet{Rahmati2016} used
  the EAGLE simulation suite to compare their simulated column density
  distribution functions (CDDFs)
of a variety of elements in various ionisation states, including \ovi,
with observations. They find that the
shape and normalisation {of the CDDF of \ovi\,} both strongly depend on the stellar
feedback efficiency due to its effect on the amount of metals produced. But they
note that \ovi\, is directly dependent on the strength of the
feedback, presumably because the temperature structure effects how
much \ovi\, is collisionally ionised}

Concentrating on \ovi, \citet{Suresh2015} make a detailed study of the offset between the CGM from
the {\it Illustris} simulation and observations.  
In addition to the nearly order of magnitude offset in column densities,
they note that their simulations show a strong correlation between
galaxy stellar mass and \ovi\, column density.  
Their galaxies also show steep column density gradients.
Additionally, they model local photoionisation to explain the low \ovi\, column
densities in simulations, but only find an effect in the inner 50 kpc
of the CGM.

{Most recently, \citet{Oppenheimer2016} tested a non-equilibrium
ionisation (NEQ) and cooling module using EAGLE zoom simulations. They find
that the NEQ does not effect the oxygen ion abundances by more than $\sim
0.1$\,dex. Their fiducial model shows a factor of two too low \ovi\,
column densities compared to observations. They argue that the NEQ
effects are strongest in 
shocks and in gas that is exposed to sources fluctuating on a short
timescale such as an AGN. }

Since S12 did well matching \ovi\, observations, another look with a newer, larger sample of galaxies is worthwhile.
The study will allow for a comparison with the trend with stellar mass that \citet{Suresh2015} found.

\subsubsection{{\sc Hi}}
{To begin to understand the large amounts of cool gas surrounding
  galaxies, \citet{Fumagalli2014} found that the \hi\, covering fractions of their simulated
  halos with  $\Mvir<10^{12}$ \Msun\, were lower than $z=2-2.5$ observations by a
  factor of $\sim30\%$. Moreover, for the halos with $\Mvir>10^{12}$
  \Msun\, they found systematically lower values (factor
  $\sim3$). However, the feedback prescription used in this work was
  not effective enough, overproducing the amount of stars and
  underestimating the amount of gas in the halos. It is worth pointing
  out that this study presented a compilation of previous results. The simulations with some of the highest covering fractions came from the {\it Eris} simulation \citep{Shen2012}, a simulation run with the same code and cooling modules we use.

Using the {\sc fire} simulation suite, \citet{FaucherGiguere2015}
also studied covering fractions of \hi. They found that the physical area the cold
gas covers stays roughly constant with cosmic time. Thus, as the
virial radius grows, the covering fraction decreases
significantly. When they compared with $z\sim2-2.5$ observations they
seem to better match the  $\Mvir<10^{12}$ \Msun\, halos but they were
still too low for the most massive systems. Originally, these authors
claimed that the reason for this could be that their simulations did
not include AGN feedback. 

In a recent study, \citet{Rahmati2015} examine the \hi\, covering
fractions using {the EAGLE simulation suite}. They
approximate the \hi\, ionisation fraction via the same fitting
functions as we do \citep{Rahmati2013} and find close agreement
between the \citet{Prochaska2013} and \citet{Rudie2012}
observations. They attribute this to their strong stellar and AGN
feedback model, although they
say that their AGN feedback prescription has a small effect. Their
comparison, however, is for larger mass halos,  $\Mvir>10^{12.5}$
\Msun, captured within their larger simulation volume with an initial gas
particle mass of $1.4\times10^6$ \Msun. 

Even more recently, \citet{FaucherGiguere2016} argue that, in fact, they can reproduce the
observed \hi\, covering fractions for the massive systems without AGN
feedback when including strong stellar winds. {Their simulations
  have a gas
particle mass of a few times $10^4 \Msun$.} This highlights the fact
that there is still no general agreement about the \hi\, covering fractions around massive high-redshift galaxies between different state-of-the-art  feedback models.}

\subsubsection{Aim of this paper}
In this paper we plan to investigate the CGM properties and
characterise the metallicity distribution and shape in Numerical
Investigation of a Hundred Astrophysical Objects, NIHAO
\citep{Wang2015} project. The NIHAO simulations are a suite of nearly 100
hydrodynamical, cosmological zoom-in galaxies across a range in mass
from $10^5 \lta \Mstar/\Msun \lta 10^{11}$. The star formation and stellar
feedback was developed in the Making Galaxies in a Cosmological
Context project \citep[MaGICC][]{Stinson2013}. 
The NIHAO project builds on this, extending the mass
range while keeping the same stellar physics at all scales. 
Despite the many orders of magnitude between our smallest and largest galaxy,
the amount of stellar mass formed in a halo with a given mass
agrees with results from abundance matching \citep{Wang2015}. 
The cold gas mass in the simulated galaxy discs also agree well with observations \citep{Stinson2015}.
It is for this reason that the NIHAO sample is an excellent
resource for exploring the physics of galaxy formation. 

Focusing on the halo mass dependence of properties of
the CGM, we will mainly look at low redshifts. 
This work is an extension of S12 to a large sample of galaxies spanning a wide range in masses.
We examine the hard-to-see low density gas surrounding galaxies,
in the ensemble of the NIHAO galaxies. 

\S2 provides an overview of the simulations used to find this result. 
\S3.1 shows the hot gas traced by \ovi\, and how the profiles compare with observations. 
\S3.2 discusses the cold gas density profiles traced by \hi\,
and takes a look at the covering fractions of \hi, comparing them to observations. 
\S3.3 studies how outflows shape different phases of the CGM. 
\S3.4 looks at the chemical composition of the CGM. 
\S4 provides context for the findings in some recent theoretical models, 
observations of gas in galaxies, and some possible implications for future observations.

\section{Simulations}
\label{sec:model}
The 86 simulations analyzed in this paper are taken from the NIHAO
project \citep{Wang2015}.  The NIHAO galaxies come from (15 $h^{-1}$
Mpc)$^3$, (20 $h^{-1}$
Mpc)$^3$ and (60 $h^{-1}$ Mpc)$^3$ cubes from \citet{Dutton2014}.
For all work done here, we use the \citet{Planck2014} cosmological parameters: $\Omega_{\rm m}$=0.3175,
$\Omega_\Lambda$=0.6825, $\Omega_{\rm b}$=0.049, $H_0=67.1$ \kms
Mpc$^{-1}$, $\sigma_8=0.8344$. The virial mass, \Mvir, of each halo is
defined as the mass of all particles within a sphere containing  $\Delta=200$ times the
cosmic critical matter density. The virial radius, \rvir, and the
virial temperature, $T_{\rm vir}$, are defined accordingly.

\subsection{ESF-Gasoline2}

We use the N-body smoothed particle hydrodynamics (SPH) solver
\textsc{gasoline} \citep{Wadsley2004} with the modified version of hydrodynamics described
in \citet{Keller2014}. Their modified version of hydrodynamics removes spurious
numerical surface tension and improves multiphase mixing. This arises
from the $P/\rho^2$ calculation as proposed by
\citet{Ritchie2001}, although we do not use densities calculated using equal pressure.

We refer to our version as \gastwo. Instead of the \citet{Keller2014} superbubble
feedback, we use the ``early stellar feedback'' ({\sc esf}) model that is described below. 

Our simulations use the metal diffusion as described in
\citet{Wadsley2008}, but we do not use the diffusion of thermal energy
between particles, because it is incompatible with the blastwave
feedback that delays cooling.

\gastwo\, also uses the timestep limiter \citep{Saitoh2009} which
allows cool particles to behave correctly when a hot blastwave hits them.
\gastwo\, uses the Wendland $C2$ function
for its smoothing kernel \citep{Dehnen2012}, helping to avoid pair
instabilities. The treatment of
artificial viscosity has been modified to use the signal velocity as
in \citet{Price2008}.  

The cooling is as described in \citet{Shen2010} and was calculated
using \textsc{cloudy} (version 07.02; \citet{Ferland1998}) tables that
include
photoionization and heating from the \citet{Haardt2005} UV background, Compton cooling, and hydrogen, helium and metal cooling from 10 to $10^9$ K.  In the dense, interstellar medium gas we do not impose any shielding from the extragalactic UV field as the extragalactic field is a reasonable approximation in the interstellar medium.  

\begin{figure}
  \centering
 \includegraphics[width=0.5\textwidth]{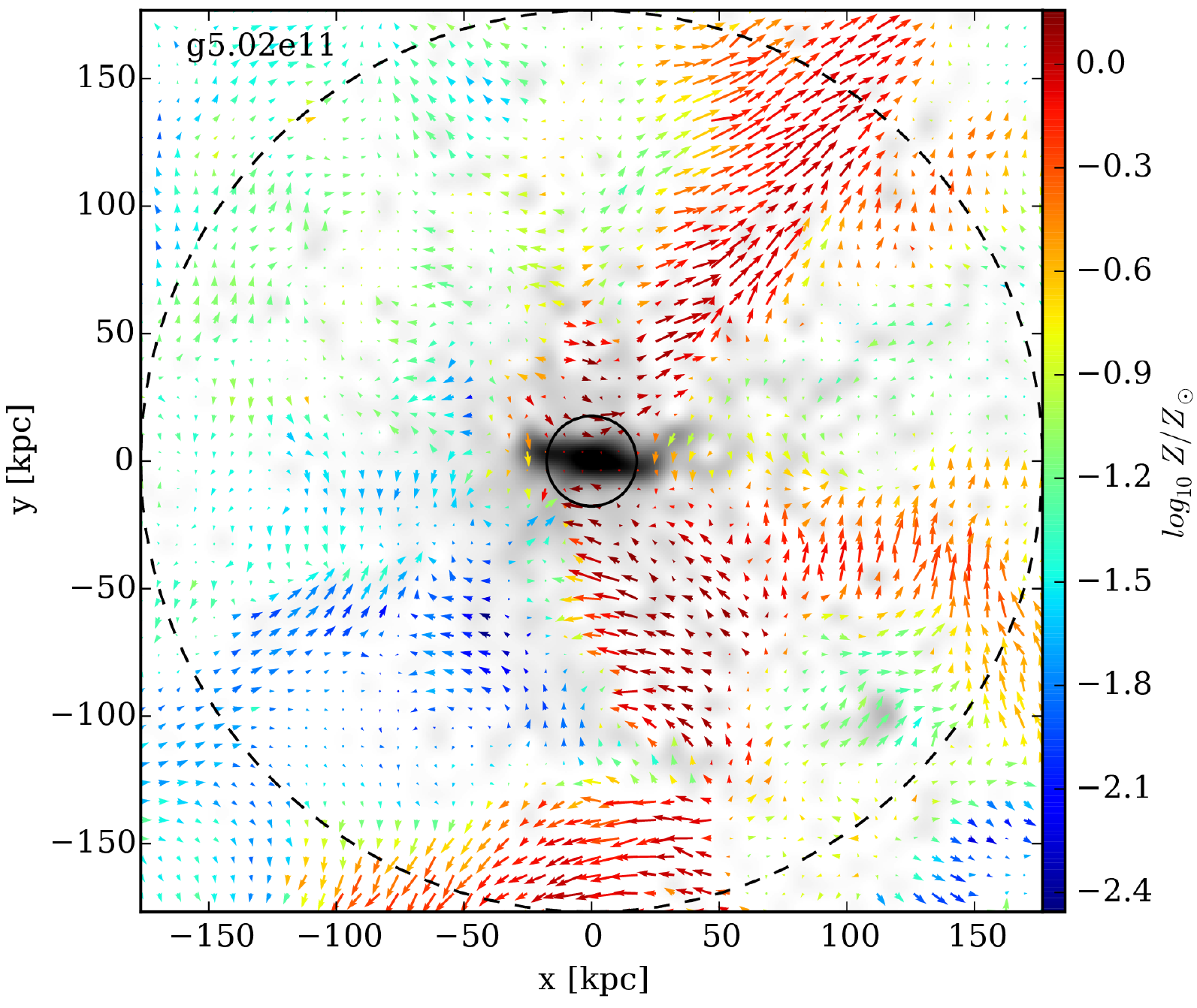}
  \caption{Edge on view of the velocity field in slice
    \rvir$\times$\rvir$\times2\,$kpc for one
    NIHAO galaxy at $z=0$, g5.02e11. Colors
  indicate the average metallicity, $\log Z/$Z$_{\odot}$, while the grey shaded areas show
  mean gas density in the slice. The virial radius and 10\% of it are shown by the
  dashed and solid circles, respectively. Metal-rich inflows and outflows
  can be traced approximately perpendicular to the plane of the disc.}
  \label{fig:vfield}
\end{figure}

\subsection{Star Formation and Feedback}
\label{sec:feedback}

The NIHAO simulations use the star formation recipe described in \citet{Stinson2006} that we summarize here.  Stars form from cool ($T < 15,000$ K), dense gas.  The metal cooling readily produces dense gas, so the star formation density threshold is set to the maximum density at which gravitational instabilities can be resolved, $\frac{50 m_{\rm gas}}{\epsilon_{\rm gas}^3}$($n_{th} > 10$ cm$^{-3}$), where $m_{\rm gas}$ is the gas particle mass and $\epsilon_{\rm gas}$ is the gravitational softening of the gas.  Such gas is converted to stars according to the equation

\begin{equation}
\frac{\Delta M_\star}{\Delta t} = c_\star \frac{M_{\rm gas}}{t_{\rm dyn}} .
\end{equation}
Here, $\Delta M_\star$ is the mass of the star particle formed, $\Delta t$ is the timestep between star formation events, $8\times10^5$ yr in these simulations, and $t_{\rm dyn}$ is the gas particle's dynamical time.  $c_\star$ is the efficiency of star formation, i.e. the fraction of gas that will be converted into stars during $t_{\rm dyn}$. 

The simulations use the same stellar feedback as described in
\citet{Stinson2013}.  In this scheme, the stars feed energy back in
two epochs.  The first epoch, ``pre-SN feedback'' ({\sc ESF}), happens
before any supernovae explode.  It represents stellar winds and
photoionization from the bright young stars, and {the efficiency
parameter} is set to $\epsilon_{ESF}$=13\%.  Radiative cooling is left {\it on} for the pre-SN feedback.
 
The second epoch starts 4 Myr after the star forms, when the first supernovae start exploding.  Only supernova energy is considered as feedback in this second epoch.  
Stars $8$ M$_\odot <$ M$_\star < 40$ M$_\odot$ eject both energy and
metals into the interstellar medium gas surrounding the region where
they formed.  Supernova feedback is implemented using the blastwave
formalism described in \citet{Stinson2006}.  Since the gas receiving
the energy is dense, it would quickly be radiated away due to its
efficient cooling.  For this reason, cooling is delayed for particles
inside the blast region for $30$ Myr. {This time is extended if a
particle resides in the blast region of multiple supernovae.}

 \begin{figure*}{!htbp}
  \centering
 \includegraphics[width=\textwidth]{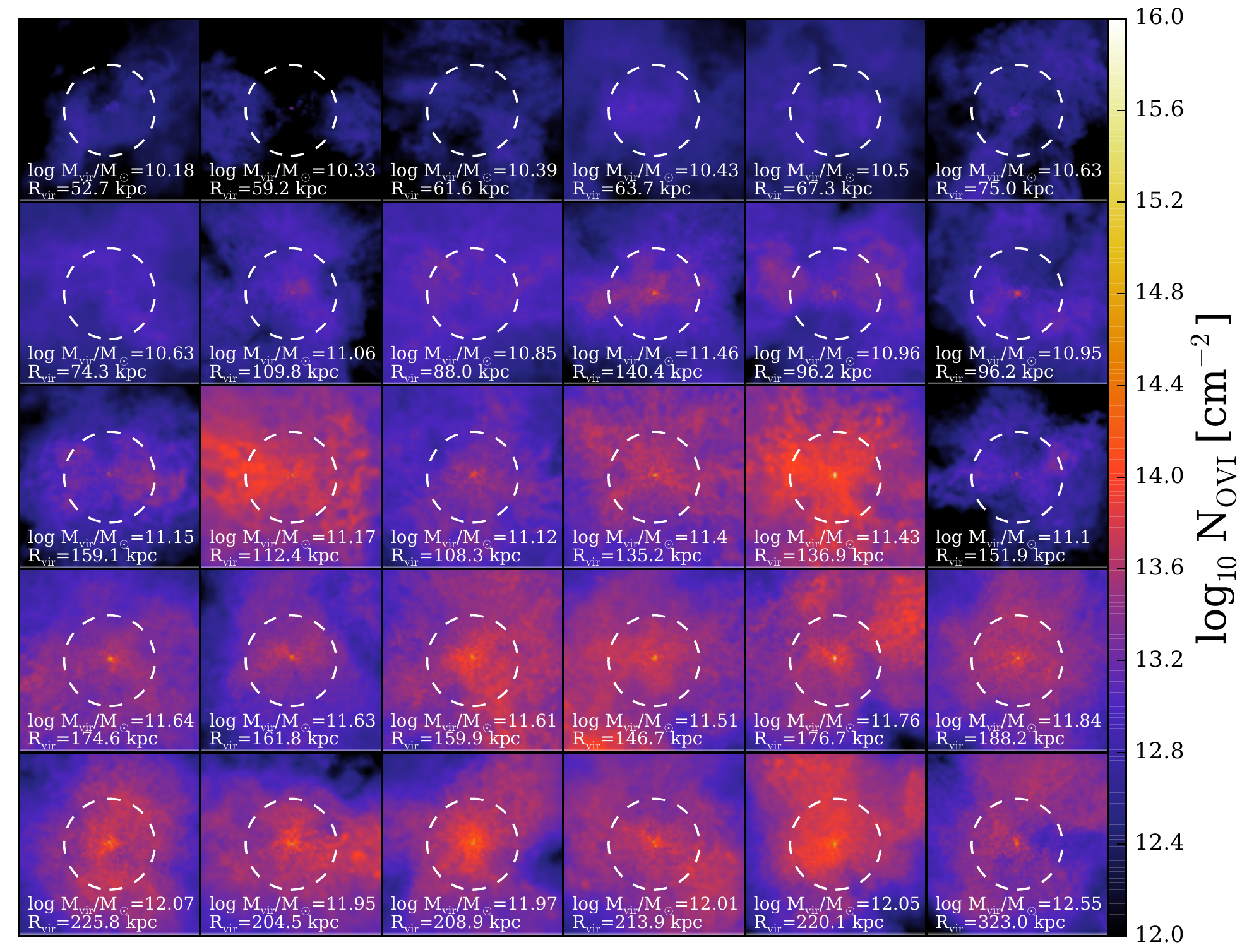}
  \caption{\ovi\, column density maps for the same 30 NIHAO galaxies as in
    \citet{Wang2015}.  The CGM images are oriented such that the discs are face-on. Each image is $2\rvir$ on each side,
    while the dashed white circle indicates the virial radius.}
  \label{fig:ovimap}
\end{figure*}

\subsection{Calculating \hi\, fractions}
\label{sec:rahmati}
{Unlike S12, which used an optically thin {\sc Cloudy} calculation to
compute the \hi\, neutral fraction, we use an analytic fits to radiative transfer calculations
presented in \citet{Rahmati2013}. 
\citet{Rahmati2013} post-processed SPH simulation using the radiative
transfer code {\sc Traphic} detailed in \citet{Pawlik2008} and
\citet{Pawlik2011}. They show that the fitting functions, which
include self-shielding density thesholds dependent on density and
redshift, produce
results in excellent agreement with their full radiative transfer calculations.}

The {\sc Cloudy} calculation used in S12 found hydrogen ionization fractions 
as high as 0.3 even at densities above 0.1 cm$^{-3}$. 

Table 2 in \citet{Rahmati2013} presents self-shielding density thresholds at
discrete redshifts and for different cosmic ionization backgrounds.  
We calculate the self-shielding density threshold by interpolating between 
the redshift values for the \citet{Haardt2001} cosmic ionizing UV background. 
The \citet{Rahmati2013} density threshold rises from $1.1\times10^{-3}$ cm$^{-3}$ at
$z=0$ to $8.7\times10^{-3}$ at $z=2$ before decreasing to higher
redshifts.

\section{Results}
\label{sec:obs}
Our study of the CGM starts with a visual inspection of warm and cool gas 
in ionization lines that are commonly observed, {\sc Ovi} and {\sc Hi}.  
For the higher mass galaxies, we compare these with observations.  
{Then we} investigate trends of the shape and extent of the CGM as a function of total galaxy mass.

In Fig. \ref{fig:vfield} we show a $z=0$ slice of one (arbitrary) NIHAO galaxy
viewed side-on (thickness $=2\,$kpc) {to give us a qualitative
impression of gas flows in the halo.} The gray shading represents gas density inside the slice. We
 overlay the gas velocity field directions as arrows. Arrow length
 scales with velocity and arrow color shows the average metallicity of
 the gas in units of solar metallicity. The dashed black line
 encompasses the virial radius. Large, metal-rich ($\log
 Z$/Z$_{\odot}>-0.4$) inflows are visible in the lower right hand corner,
 while at the same time metal-rich outflows are being driven in the
 upper right hand corner.

\subsection{Hot gas}

\begin{figure*}
  \centering
\includegraphics[width=0.80\textwidth]{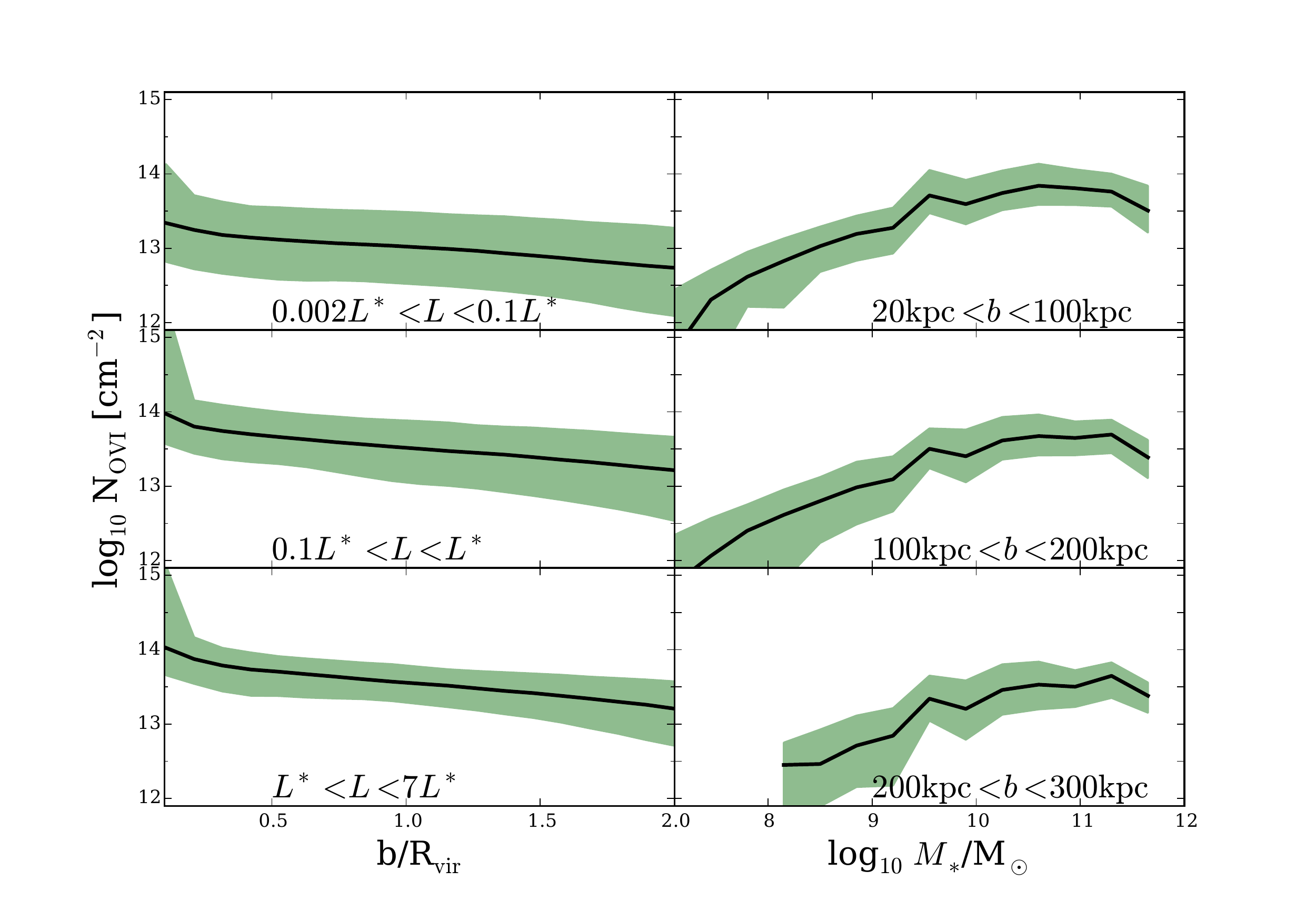}
  \caption{\ovi\, column density ($z=0$) as a function of:
impact parameter normalized to the virial radius (left panels)
and 
stellar mass (right panels).  
  All sight lines from face-on and edge-on for every NIHAO galaxy are included.
  The black solid line shows the median column density at $z=0$, while
  the green band shows the 5-95 percentile range.
  In the left panels, the column densities are divided between three luminosity bins: 
    $L<0.1L^\star$ (upper panel), $0.1L^\star<L<1L^\star$ (middle panel) and  
    $L>1L^\star$ (lower panel).  
    In the right panels, the column densities are divided between radial bins:
    $20<b/{\rm kpc}<100$ (upper panel), $100<b/{\rm kpc}<200$ (middle
    panel) and $b>200\;$kpc (lower panel).}
\label{fig:mstar_ovi}
 \includegraphics[width=0.80\textwidth]{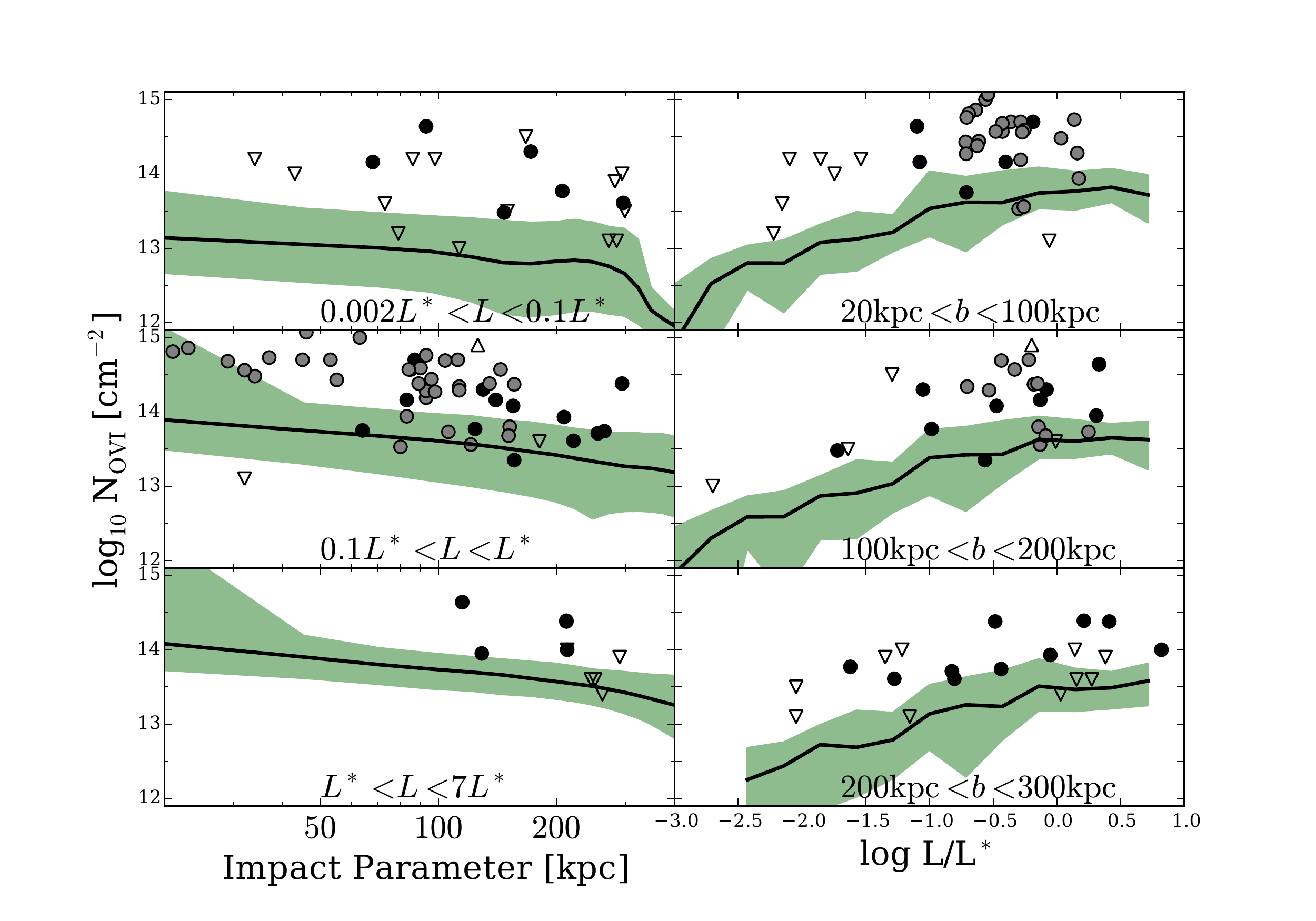}
  \caption{\ovi\, column density ($z=0$) as a function of:
impact parameter (left panels) and 
  luminosity (right panels).  The simulations and binning are the same
  as in Fig. \ref{fig:mstar_ovi} above. Observational data from \citet{Prochaska2011} are shown as black circles. 
  \citet{Tumlinson2013} data is shown as gray circles.
  Non-detected upper and optically thick lower limits are shown as 
  upward and downward facing open triangles, respectively. 
Observational points are only shown in the lower
  panels, since the virial radii and stellar masses of the
  observations are not known.}
  \label{fig:lstar_ovi}
\end{figure*}

Although this analysis uses 86
galaxies of the NIHAO suite,
Fig. \ref{fig:ovimap} shows column density maps of \ovi\, for the same
30 NIHAO galaxies shown in \citet{Wang2015}. 
The \ovi\, mass is calculated per particle as in S12 
using 1 cm thick gas slabs in Cloudy (v10.0 last described in \citet{Ferland1998}).  
The gas is radiated with a \citet{Haardt2005} UV radiation field evaluated at $z=0$.
The resulting \ovi\, mass is smoothed using the smoothing kernel and then the images 
are created as projections of the \ovi\, mass through the entire zoom
region, a $\sim4$\,Mpc
line-of-sight. 

The 30 {\sc Ovi} maps in Fig. \ref{fig:ovimap} are ordered by \mvir~and range 
from $\approx 10^{10} - 10^{12}$ \Msun. 
The x and y axes are scaled to show the gas halos out to $2\rvir$. 
Qualitatively, Fig. \ref{fig:ovimap} shows the {\sc Ovi} column and
its extent scales with \mvir~: 
the small galaxies are practically devoid of \ovi, 
while more massive galaxies show observable \ovi\, column densities out past $2\rvir$. 

To compare the galaxies with observations, we use the column density maps as
though each position was a QSO sightline to make a sample of thousands of sightlines.

Since the observed samples necessarily combine observations from a
variety of galaxies, {Fig. \ref{fig:mstar_ovi}  and Fig. \ref{fig:lstar_ovi} show
combinations of column density profiles of all 86 simulated galaxies. }
All the sightlines from each group of simulations are combined. 
{Fig. \ref{fig:mstar_ovi} shows the median \ovi\, column density as a function of impact parameter {$b$} normalised to each
respective virial radius (left panels).} The impact parameter directly
is shown in the left panels of Fig. \ref{fig:lstar_ovi} as the black solid line.
The green-gray shaded area shows the 5-95 percentile range of our data. 
Individual galaxy profiles are shown in fig. \ref{fig:oviprof} of the
appendix for the same 30 galaxies as Fig. \ref{fig:ovimap}.
{The functional form of the combined profiles in Fig. \ref{fig:lstar_ovi} is very
  similar to the individual galaxy profiles seen in
  Fig. \ref{fig:oviprof}. This indicates that the galaxy-to-galaxy
  variations are small and that the combined trends mainly show the
  variation in profiles. Solely the galaxies with the smallest
  luminosities ($L<0.002L^{*}$, the profiles without observational
  data in Fig. \ref{fig:oviprof}) show a larger galaxy-to-galaxy
  scatter. These are neither compared to observations, nor included in
  Fig. \ref{fig:lstar_ovi}.}

Observed data from galaxies of similar luminosities are overlaid on
top of the simulated columns of Fig. \ref{fig:lstar_ovi},  
according to the luminosity ranges used in \citet{Prochaska2011},  
$0.01L^\star<L<0.1L^\star$, $0.1L^\star<L<1L^\star$ and $L>1L^\star$.
In the luminosity range $0.1L^\star<L<1L^\star$, data from \citet{Tumlinson2011} as gray circles. 
Luminosities are calculated for the simulated galaxies using Padova simple
stellar populations \citep{Marigo2008, Girardi2010}.
L$^{\star}$ is taken as M$_V=-21.12$ as reported in \citet{Prochaska2011}.
\citet{Tumlinson2011} gives SDSS $r$-band galaxy magnitudes; we use
M$_r=-20$ for L$^\star$.
While the COS-Dwarfs project \citep{Bordoloi2014} observed galaxies of similar
luminosity as our lowest mass simulations, they have yet to publish \ovi\, columns. 

When exact data values are given, they are shown as filled black circles. 
Upper limits are shown as downward pointing triangles, 
and lower limits are shown as upward pointing triangles. 

{No observations are shown in Fig. \ref{fig:mstar_ovi} since neither
the virial radii nor the stellar masses are available for the observed galaxies.}

The general trend of increasing column density with increasing $M_{\rm star}$
(so also total luminosity) is apparent. 
For this analysis, we will focus on the $0.1<L/L^\star<1$ range, 
since many of the observations in the lower luminosity bins are only upper limits, 
and the sample size is mainly limited in the higher luminosity bin. 
In the $0.1<L/L^\star<1$ range, the simulations are about half a dex lower than the median
of the observations.
Such an offset is typical of comparisons of \ovi\, observations with
simulations \citep{Ford2015, Hummels2013,Suresh2015}. {Various causes are
mentioned, among them that the oxygen, which seems available in
sufficient amounts, is not in the correct
ionisation state. Assuming collisional ionisation dominates, this
points to a problem with the thermal structure of the CGM produced in
hydrodynamical simulations. Further discussion of this in \S4.}

There are two significant differences that could explain the
apparent discrepancy between the NIHAO sample and
analysis of S12.
First, the NIHAO sample is much larger than the 2 galaxies shown in S12.
In S12, only the brighter galaxy well matched the observations.  

\begin{figure}
  \centering
 \includegraphics[width=0.5\textwidth]{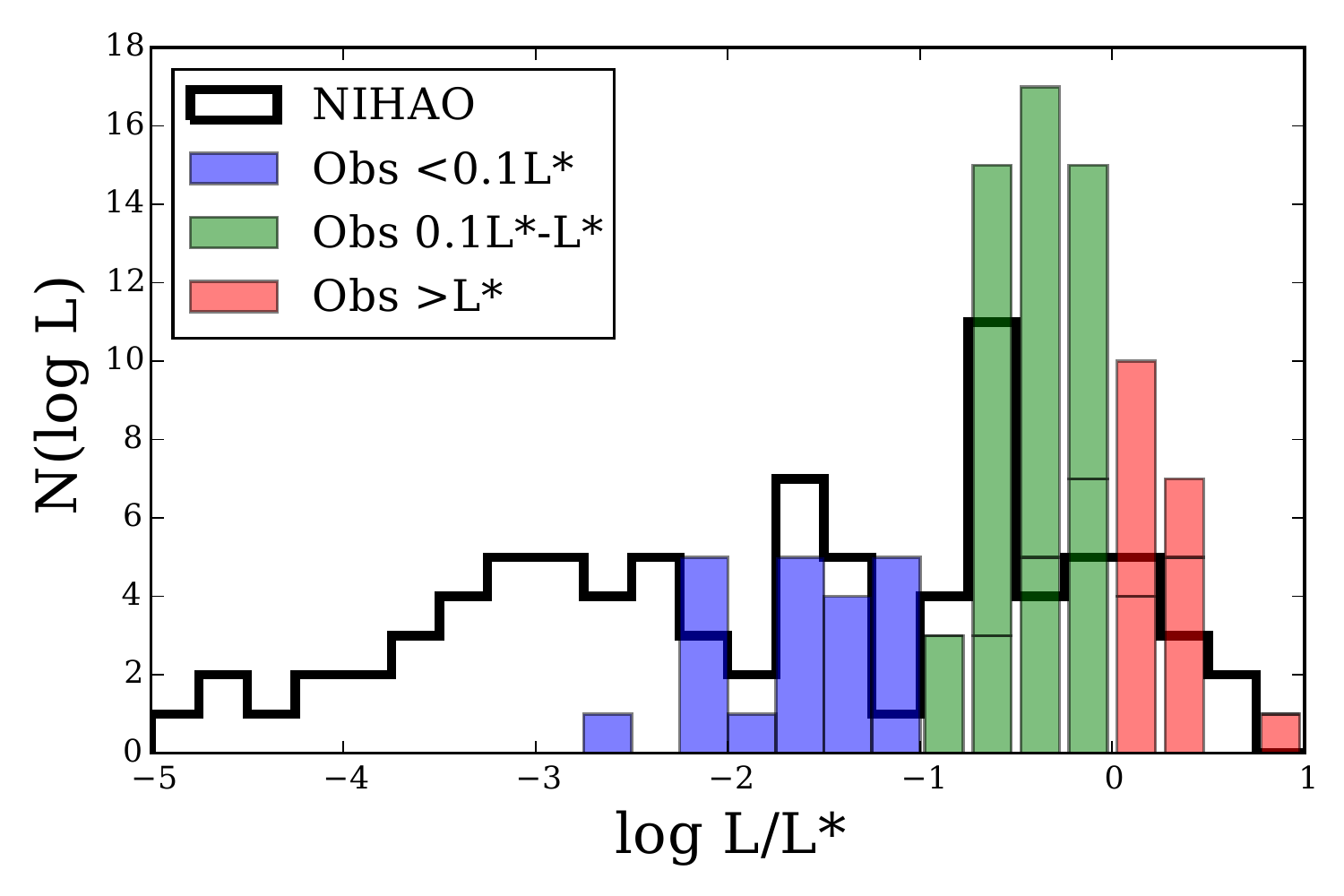}
  \caption{Luminosity distribution of NIHAO galaxies is presented as
    a black line. The
    coloured bars (blue, green, red) show the distribution of
    observations from \citet{Prochaska2011} and \citet{Tumlinson2011}.}
  \label{fig:lumhist}
\end{figure}

Fig. \ref{fig:lumhist} shows that NIHAO samples the lower half of the $0.1L^\star<L<1L^\star$ bin 
more than the \ovi\, observations, which are mostly in the upper half of the bin.
The right panels of Fig. \ref{fig:lstar_ovi} are helpful to make this comparison.
They show that at common projection radii and galaxy luminosity, 
NIHAO galaxies indeed have lower \ovi\, column densities.

Another difference between NIHAO and S12 is the stellar feedback.
S12 used a slightly different version of hydrodynamics that required
stronger stellar feedback to reduce star formation enough to fit on 
the stellar mass--halo mass relation.  
S12 and \citet{Hummels2013} showed that more stellar feedback add \ovi\, to the CGM.  
We discuss in \S \ref{sec:discussion} various solutions to this problem.
While the discrepancy is unwelcome, it is not so large that it precludes drawing
other useful information about the properties of the CGM in simulations.

For example, the \ovi\, column density profiles in the left panel of Fig. \ref{fig:lstar_ovi} are remarkably flat.
From the centre out to 250 kpc, N$_{\rm OVI}$
varies less than 0.2 dex in the $0.002<L/L^\star<0.1$ range.
In both $0.1<L/L^\star<1$ and $1<L/L^\star<10$ luminosities, 
N$_{\rm OVI}$ slowly, but steadily declines 0.7 dex over 300 kpc.
S12 and \citet{Hummels2013} both showed that with
lower feedback, the \ovi\, does not extend to high radii.  
However, with sufficient feedback to match the stellar mass a galaxy should have, 
the \ovi\, extends in a flat distribution as far as the observations.
\citet{Suresh2015} shows a much steeper profile, close to a 2 dex 
decline in N$_{\rm OVI}$ over 300 kpc in $Illustris$.
\citet{Ford2015} shows little decline to 150
kpc in either their energy dependent or constant wind model. But
we note that the \citet{Ford2015} analysis compares equivalent widths
instead of column densities.

{To examine the question of how much \ovi\, depends on stellar mass or luminosity,
the right panels of Fig. \ref{fig:mstar_ovi} and Fig. \ref{fig:lstar_ovi} show column density 
as a function of stellar mass and galaxy luminosity, respectively, binned into three ranges of the}
{\it impact
parameter} $b$: 
$20<b/{\rm kpc}<100$, $100<b/{\rm kpc}<200$ and $b/{\rm kpc}>200$. 
The median column density is shown as the black line. 
The green area outlines the spread between the 5th and 95th percentiles in the 
0.25 dex luminosity bins.
The simulations follow the general trends of the observations, albeit
nearly a dex below the
observations in all radial ranges. 

{The simulations clearly show a correlation between stellar mass
  and N$_{\rm OVI}$ in Fig. \ref{fig:mstar_ovi}. This relation is
  mirrored in the luminosity trends in Fig. \ref{fig:lstar_ovi}.}
N$_{\rm OVI}$ rises steadily from unobservable to a threshold N$_{\rm OVI}$
around $10^{14}$ cm$^{-2}$.  
The simulations reach the $10^{14}$ cm$^{-2}$ threshold at higher galaxy
luminosities the higher the impact parameter range is, from $10^{-0.5} L^\star$ in the central bin,
to $L^\star$ in the medium distance radius to $>L^\star$ for the
largest values of $b$.
It makes sense that brighter galaxies that formed more stars have more
enriched material further away.
The issue with the simulations seems to be that the threshold N$_{\rm OVI}$
is lower than what is observed.

{N$_{\rm OVI}$ observations are often interpreted as showing little
  to no dependence on luminosity.
More observations are necessary to make definite statements, but the
right panels of Fig. \ref{fig:lstar_ovi} hint at a correlation. To
distinguish whether there is a trend or not, more observations are
necessary at lower luminosities, since there are very few \ovi\, detections
below 0.1$L^\star$.  
The trend in the simulations is visible because they predict N$_{\rm OVI}$ below the detectable limit.}

\begin{figure*}
  \centering
 \includegraphics[width=\textwidth]{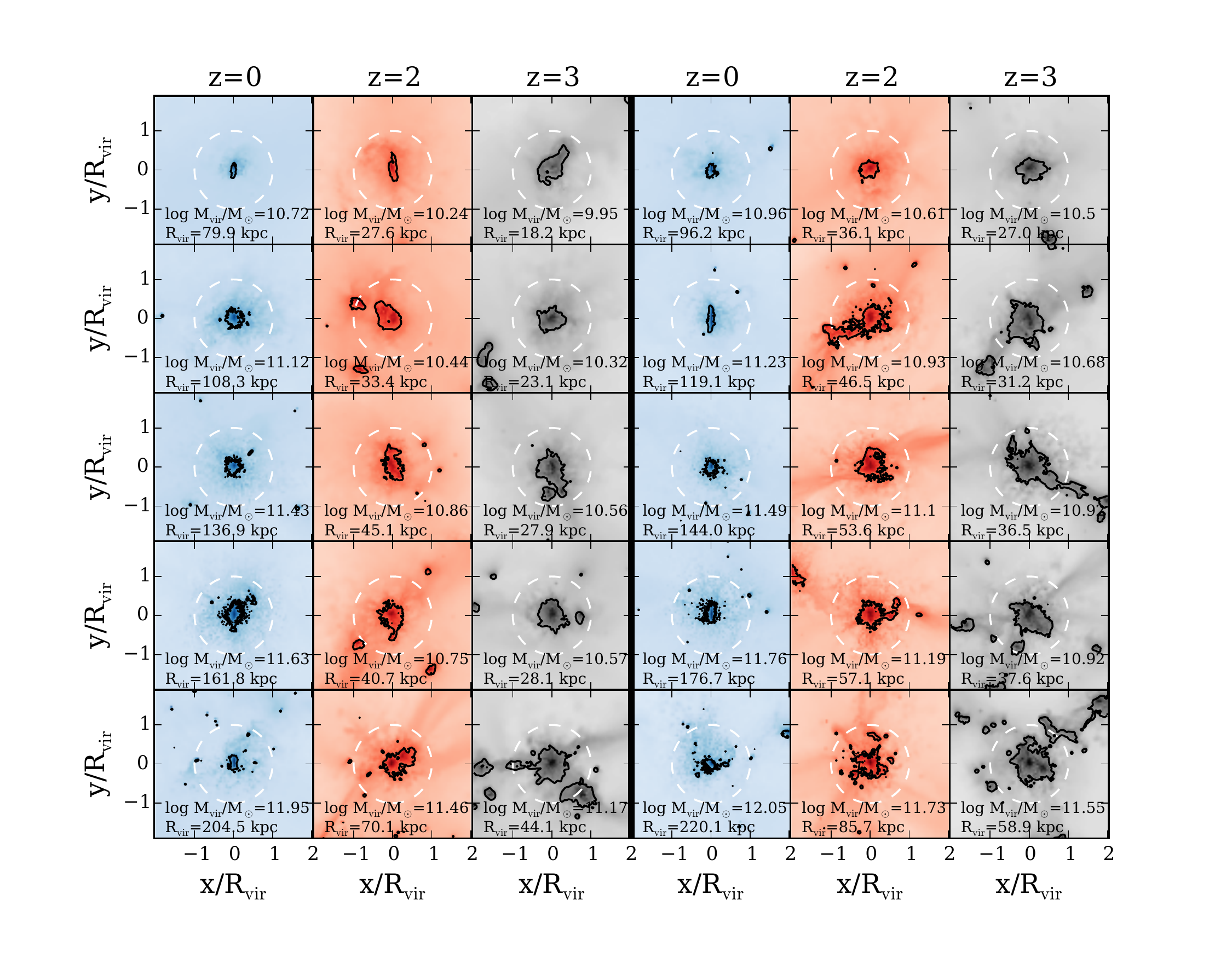}
  \caption{Face-on \hi\, column density maps for 10 NIHAO galaxies. 
  The first and fourth column shows $z=0$,
    while the two neighbouring columns show $z=2$ and $z=3$,
    respectively. The black contours indicate the $\log N_{HI}>17.2$ cm$^{-2}$
    threshold, above which the gas is considered optically thick. Each
    box is $2 \rvir$ on each side,
    while the dashed white circle indicates the virial radius.}
  \label{fig:himap}
\end{figure*}

\subsection{Cold gas}

\begin{figure*}
  \centering
 \includegraphics[width=0.95\textwidth]{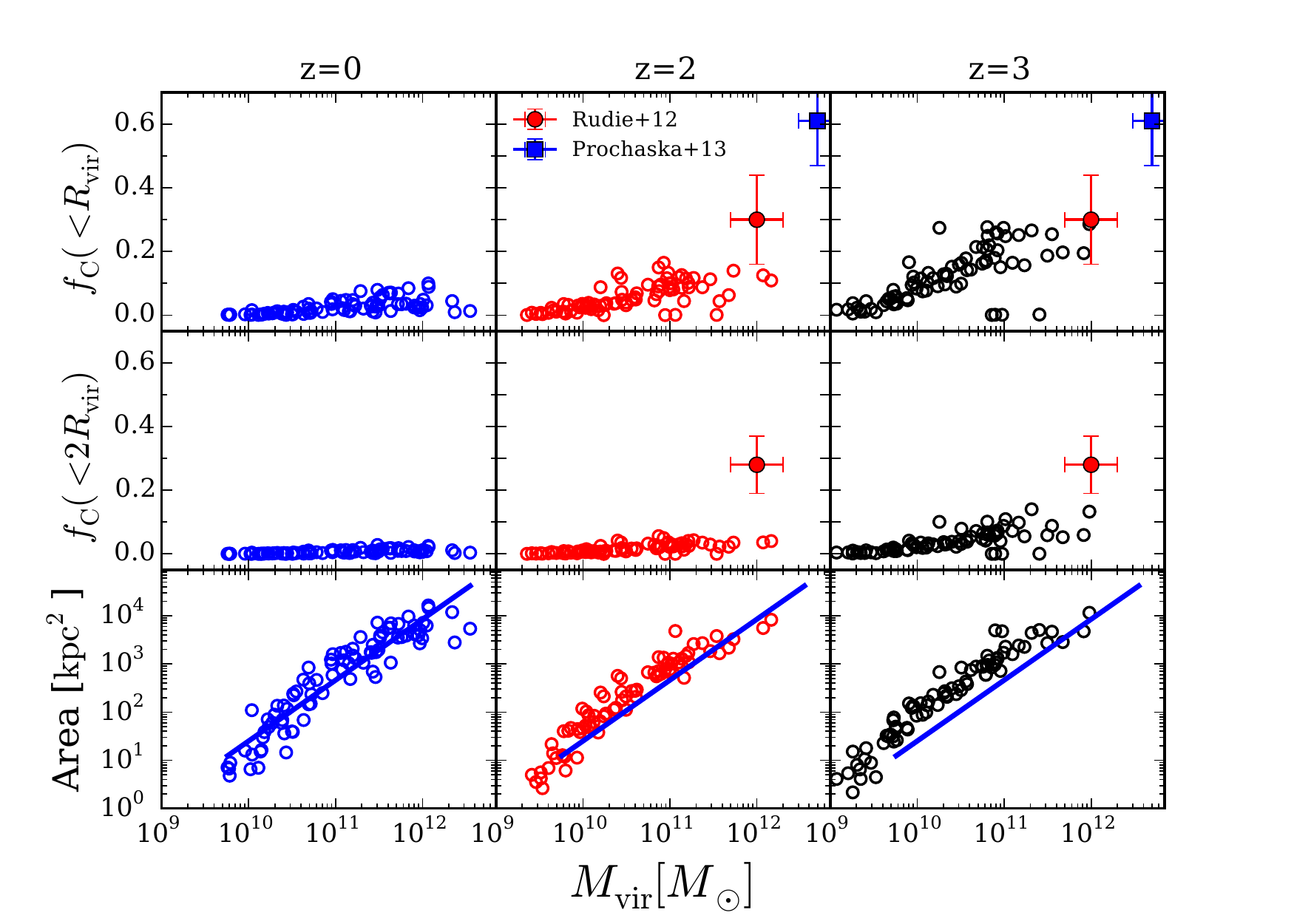}
  \caption{Fraction of area covered by \hi\, with a column density for
    which $\log N_{HI}>17.2$ cm$^{-2}$. Upper panels show the projected area
    fraction inside \rvir\,for $z=0, 2$ and $3$. Middle panels show the
    covering fraction in 2\rvir. The red error bars are from
    observations by \citet{Rudie2012}, who observed LBGs in a redshift
    range $z=2-2.5$. The blue error bars are from \citet{Prochaska2013}
    measurements of Lyman limit systems (LLS) via luminous
    quasi-stellar objects (QSOs). We plot the same observational points on our $z=2$
    and $z=3$ panels. The middle panels also include an observational
    point from \citet{Rudie2012} for the fraction inside 2\rvir. The
    lower panels show the area covered (in kpc$^2$). The blue line is
    a power law fit to the $z=0$ data and overplotted on the $z=2$ and
    $z=3$ panels for reference.
  }
  \label{fig:covfrac}
\end{figure*}

\begin{figure*}
  \centering
\includegraphics[width=0.80\textwidth]{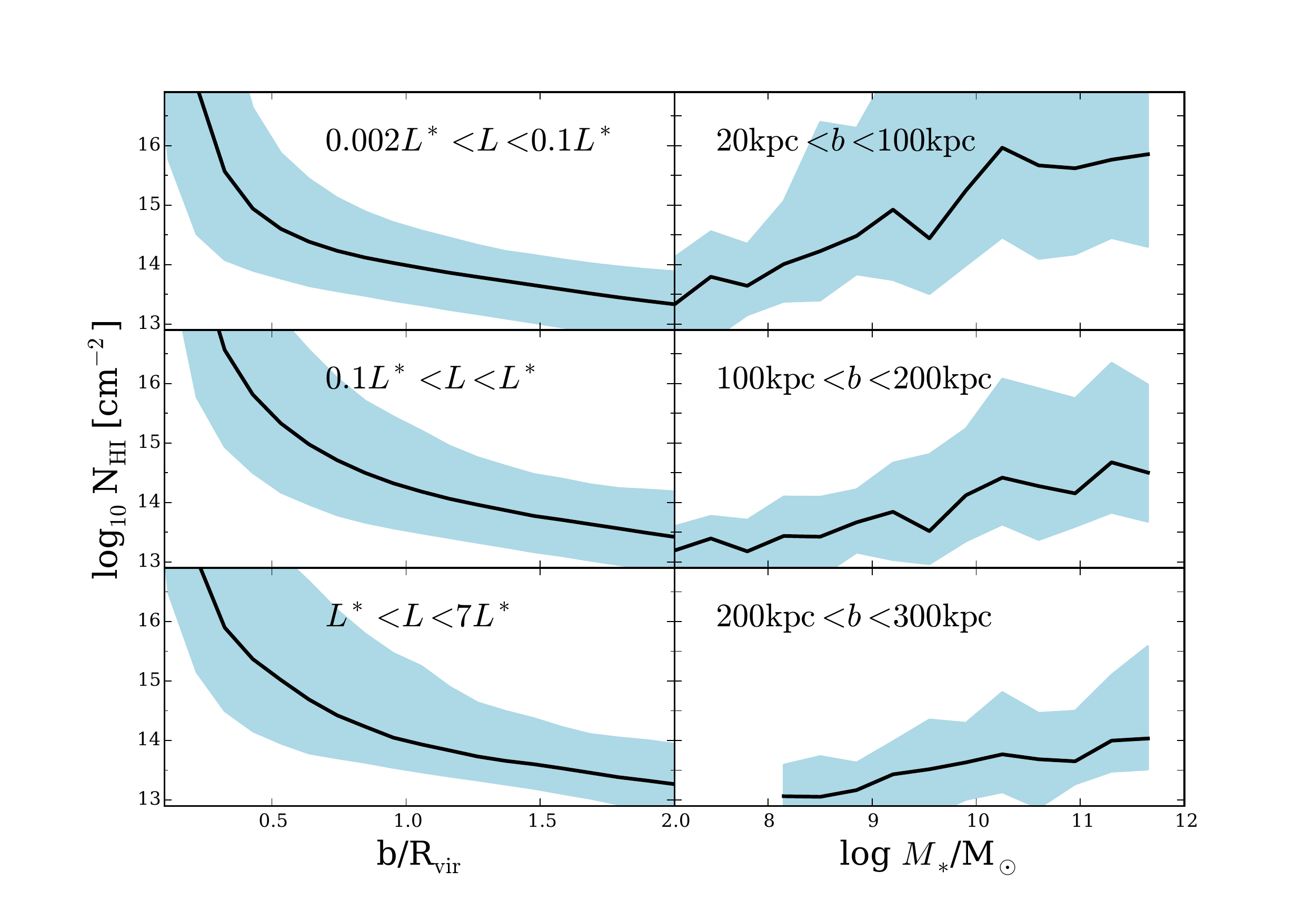}
  \caption{\hi\, column density ($z=0$) as a function of:
impact parameter normalized to the virial radius (left panels)
and 
stellar mass (right panels).  
  All sight lines from face-on and edge-on for every NIHAO galaxy are included.
  The black solid line shows the median column density at $z=0$, while
  the green band shows the 5-95 percentile range.
  In the left panels, the column densities are divided between three luminosity bins: 
    $L<0.1L^\star$ (upper panel), $0.1L^\star<L<1L^\star$ (middle panel) and  
    $L>1L^\star$ (lower panel).  
    In the right panels, the column densities are divided between radial bins:
    $20<b/{\rm kpc}<100$ (upper panel), $100<b/{\rm kpc}<200$ (middle
    panel) and $b>200\;$kpc (lower panel).}
\label{fig:mstar_hi}
 \includegraphics[width=0.80\textwidth]{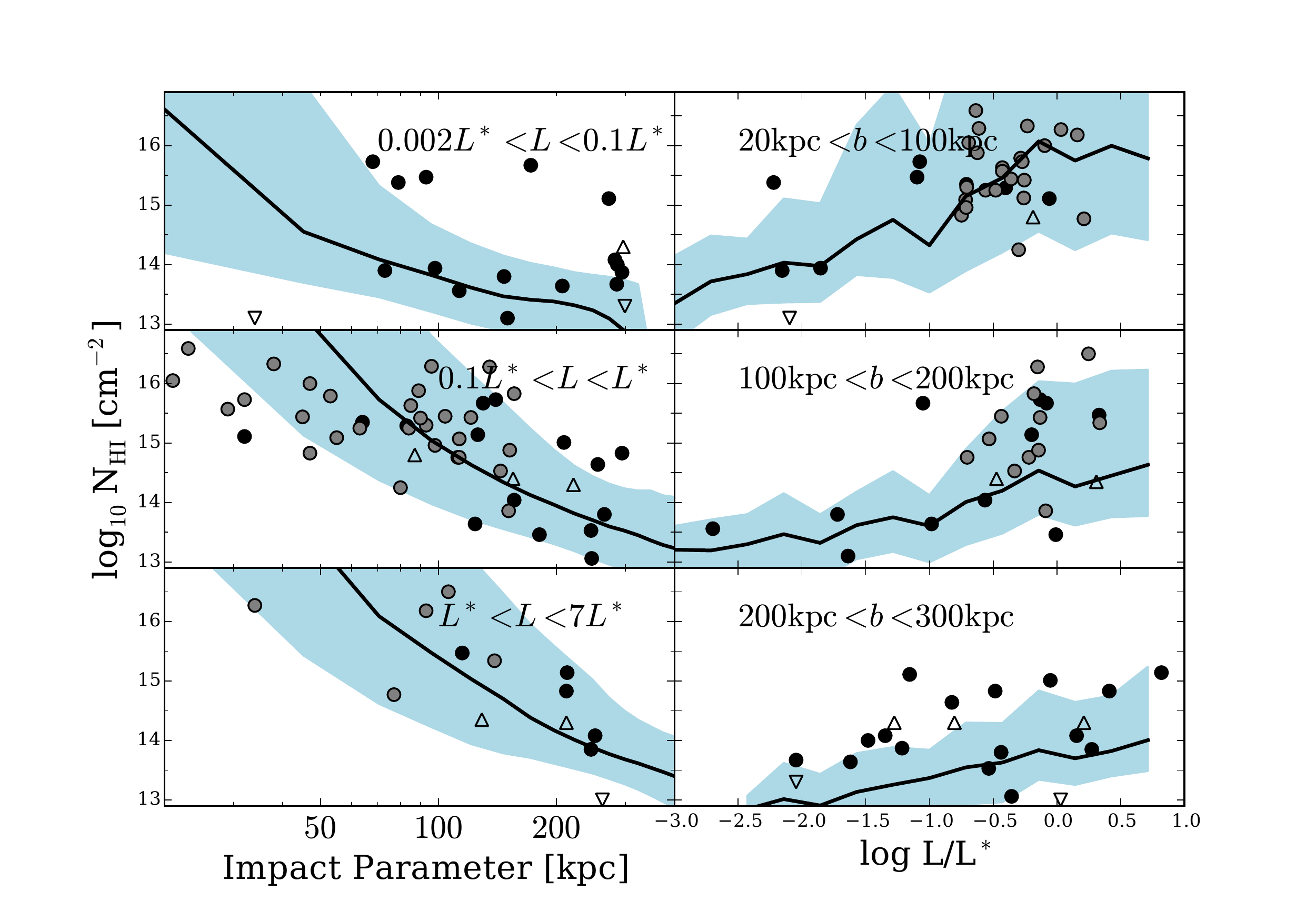}
  \caption{\hi\, column density ($z=0$) as a function of:
impact parameter (left panels) and 
  luminosity (right panels).  The simulations and binning are the same
  as in Fig. \ref{fig:mstar_ovi} above. Observational data from \citet{Prochaska2011} are shown as black circles. 
  \citet{Tumlinson2013} data is shown as gray circles.
  Non-detected upper and optically thick lower limits are shown as 
  upward and downward facing open triangles, respectively. 
Observational points are only shown in the lower
  panels, since the virial radii and stellar masses of the
  observations are not known.}
  \label{fig:r_hi}
\end{figure*}

Quasar absorption line spectra also include many low ionization
absorption lines that indicate the presence of cool gas.  
Here, we examine \hi\, as the proxy for cool gas in the simulations.  
Since hydrogen is so abundant, getting the correct ionization fraction requires
following radiative transfer.  
\citet{Rahmati2013} found that a local self-shielding approximation gives results reasonably
consistent with the full radiative transfer solution.  
We report \hi\, column densities based on \citet{Rahmati2013} and note that this is a
different treatment than that used in S12, which used an equilibrium
CLOUDY model for determining \hi\, fractions. 
Details can be found in \S \ref{sec:rahmati}.

Fig. \ref{fig:himap} shows \hi\, maps at $z=0$ (blue), $z=2$ (red), and $z=3$ (grey) 
for a representative selection of 10 galaxies spanning the NIHAO mass range.  
Each image is scaled to be 4 \rvir\,across {and along a $\sim4\,$Mpc
  line-of-sight (A discussion of this is in section \ref{sec:discussion}).}
We include higher redshift analysis because some of the most surprising results about the CGM 
are the high covering fractions of low ionisation lines.  
Such values have been reported at $z>2$ \citep{Rudie2013, Prochaska2014, Crighton2015}.

\begin{table*}
\caption{Covering fractions of \hi\, for halos with
  $\Mhalo\sim 10^{12}\Msun$ at $z=2$ and $z=3$ and for
  \citet{Rudie2012}. }
\begin{tabular}[c]{l|c|c|c|c|c|c}
\hline
\hline
log$(N_{\rm HI}/({\rm cm}^{-2}))$ & z=2 & z=2 & z=3 & z=3 & Rudie+2012 & Rudie+2012\\
& $f_{\rm C}(<\rvir)$ & $f_{\rm C}(<2\rvir)$ &  $f_{\rm C}(<\rvir)$ & $f_{\rm C}(<2\rvir)$ & $f_{\rm  C}(<\rvir)$ & $f_{\rm C}(<2\rvir)$ \\
\hline 
$>15.5$& $45 \pm 3$\% &$20 \pm 1$\%& $73 \pm 7$\% & $41\pm 10$\% & $90\pm9$\% & $68\pm 9$\%\\
$>17.2$ & $12 \pm 1$\% & $3.8 \pm 0.2$\% & $24 \pm 5$\% & $10 \pm 4$\% & $30\pm14$\% & $28\pm 9$\%\\
$>19.0$ & $2.5 \pm 0.2$\% & $0.8 \pm 0.1$\% &  $5 \pm 3$\% & $2 \pm 1$\% & $10\pm9\%$ & $8\pm 5$\%\\
$>20.3$ & $0.8 \pm 0.1$\% & $0.25 \pm 0.04$\% &  $1.4 \pm 0.7 \%$ & $0.5 \pm 0.3$\% & $0^{+10}_{-0}$\% & $4\pm 4$\%\\
\hline
\end{tabular}
\label{tab:covfrac}
\end{table*}

The white dashed circle is drawn at \rvir\,and the
black contours show where $\log N_{HI}>17.2$, 
sometimes used as a limit for where \hi\, is considered optically thick. 
Log $N_{HI}>17.2$ is used as a threshold in Fig. \ref{fig:covfrac} to calculate
the covering fractions of \hi\, inside \rvir\,and 2\rvir\,for the three
redshifts shown using the entire NIHAO sample. 
Since covering fraction measurements require many quasar absorption lines, 
there are only 2 reported observational data points at $z=2-3$.
Our estimates are compared with the observations of \citet{Rudie2012}, 
who looked at a sample of Lyman Break Galaxies (LBGs) and Damped Lyman
Alphas (DLAs) in the redshift range $z=2-2.5$. 
LBGs have a higher \mvir\,than nearly every NIHAO galaxy
\citep[but for comparisons of larger galaxies simulated with a similar
feedback scheme, see][]{FaucherGiguere2015, FaucherGiguere2016}. 
We also compare to a data point of \citet{Prochaska2013}, who measured
Lyman limit systems in a redshift range $z=2-2.5$ with the aid of
luminous background quasars.

With that caveat, it appears the simulated covering fractions inside \rvir\, 
reach nearly the level of the \citet{Rudie2012} data point at $z=3$. 
However, when comparing with the larger 2\rvir\,measurement, the 
simulated covering fraction halves.  
As we will see, observations repeatedly find more cool gas distributed
far from galaxies than simulations.  
In \citet{Rudie2012}, the cool gas seems to extend well beyond \rvir.

After the galaxies have evolved to $z=2$, the simulations 
show about half the covering fractions they did at $z=3$.
So, the simulations show covering fractions less than one half of those
found in the observations.  

The lowest three panels in Fig. \ref{fig:covfrac} show that the simulations
actually maintain a constant optically thick physical area according to the galaxy's \Mvir.
The blue line in each panel shows the fit to the $z=0$ data and is repeated 
in each panel as a reference for the reader. 
From this, it can be seen that the area covered does not increase much over time and that 
the covering fractions are decreasing mostly due to the increasing \rvir\,as the
background density decreases and the Universe evolves.

{We also compare our most massive systems with further data from
  \citet{Rudie2012} in table \ref{tab:covfrac}. For each covering
  fraction at different column
  density thresholds, our $z=3$ is closer to the observations, but
  most are still too low by more than 1 sigma. {It should be noted
  that our line-of-sight is limited to the zoom region, $\sim 4$\,Mpc
  which is significantly less than in observations. This biases our
  column densities to lower values.} Since other simulations
have also under-predicted the \hi\, covering fractions, this points to
an issue with the amount of \hi\, at larger radii (but see
\citealt{FaucherGiguere2016} and \citealt{Rahmati2015} for
discussion of the effect of feedback strength on \hi\, covering fractions).}

Figs. \ref{fig:mstar_hi}  and \ref{fig:r_hi} present the combination of the
data from the entire NIHAO sample (86 galaxies {at $z=0$}) in the same manner as was done for
\ovi\, in Figs. \ref{fig:mstar_ovi} and \ref{fig:lstar_ovi}. Again, individual galaxy profiles can
be viewed in Fig. \ref{fig:hiprof} of the appendix.
In both panels of Fig. \ref{fig:r_hi}, 
the black line shows the median of the combined data and while 
the shaded region shows the range of data from the 5th to 95th percentile. 
In the left panel, column density profiles are combinations of column densities
from galaxies broken into three luminosity bins. 
The simulations show a broader scatter in \hi\, than they did for \ovi.

The simulations agree much better with the \hi\, data than
they did with the \ovi\, data.  
In all three luminosity bins, the median column density line goes through the middle of the observations.  
The simulations have a similar scatter to the observations.
One possible exception is the dwarf galaxies ($L<0.1L^\star$), for which there are
observations outside 100 kpc that show higher column densities 
than the simulated column densities. 

The right panels of Fig. \ref{fig:r_hi} show how the \hi\, column density
varies as a function of galaxy luminosity, split into three impact parameter bins:
$20<b/{\rm kpc}<100$, $100<b/{\rm kpc}<200$, and $200<b/{\rm
  kpc}<300$. 

It is clear in Fig. \ref{fig:r_hi} that the scatter of \hi\, columns is largest close to the galaxy.
The median \hi\, column density in the simulations increases 
strongly when going towards smaller projected radii. This trend goes
with galaxy luminosity.  
In the two inner $b$ bins (two upper right panels of Fig. \ref{fig:r_hi}),
the simulated median column density generally follows the observations. 
The column density does not increase as much in the outer two radial $b$ bins.
In the 100-200 kpc bin, it appears that the median observed \hi\, column density
actually increases.
Again, outside 200 kpc, the observed column densities are above the simulation. 
It seems that the simulations have trouble putting enough cool gas at large distances from galaxies.
This could be an improper treatment of the hydrodynamics or an
incorrect propulsion of the winds. {We discuss reasons for the lack of
  \hi\, at large distances in \S \ref{sec:coolgas}}.

\begin{figure}
  \centering
 \includegraphics[width=0.5\textwidth]{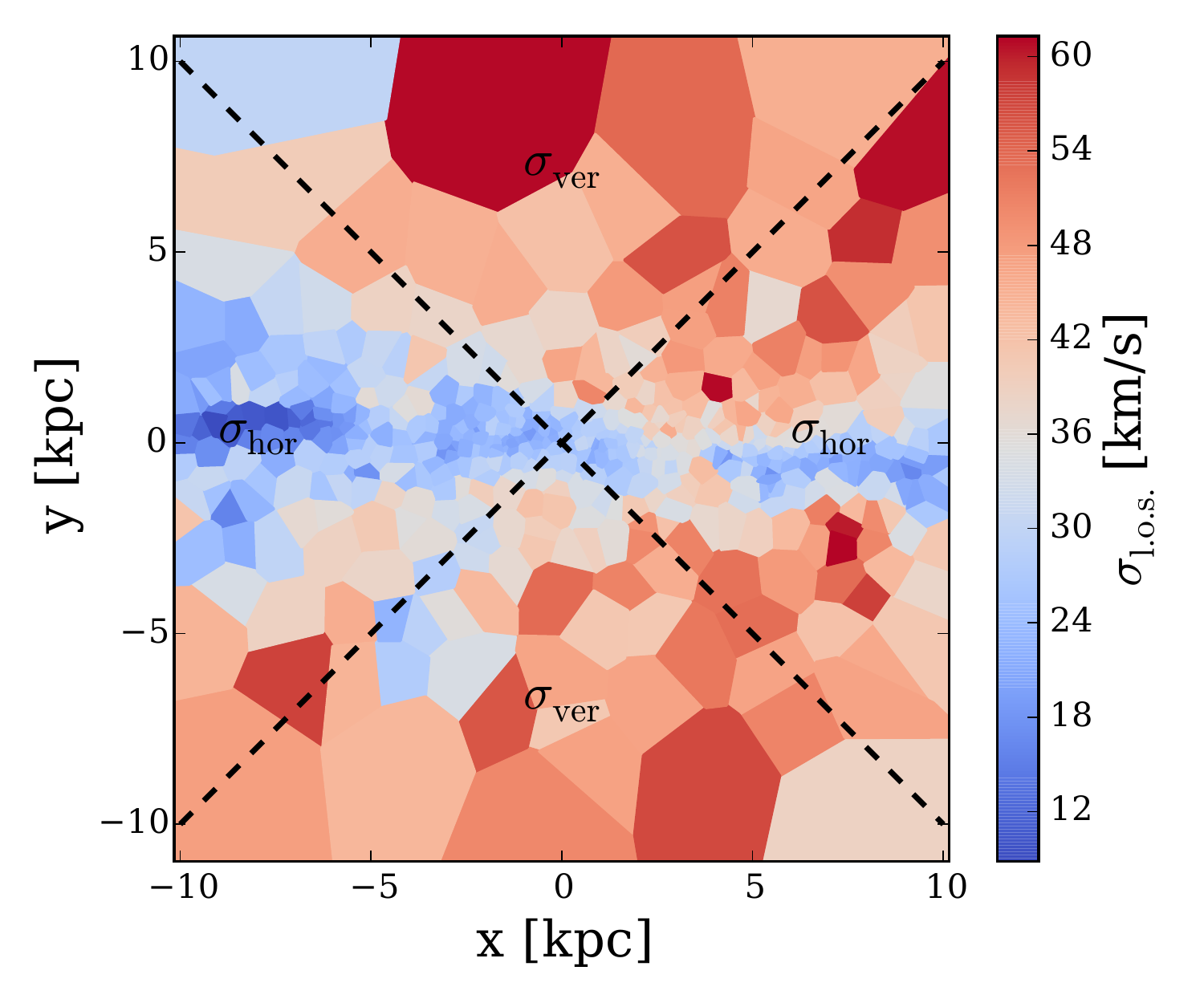}
  \caption{Edge-on velocity dispersion map of NIHAO galaxy, g5.02e11
    at $z=0$. 
  In the tessellation, the disc is apparent as lower line-of-sight (los) velocity dispersion (blue).
  Above and below the plane of the disc, the los velocity dispersion increases (red). 
  \citet{Wild2014} proposes dividing the velocity dispersion map using 
  the black dashed lines to detect a bipolar outflow. 
  $\sigma_{\mathrm{ver}}$ is the mean velocity dispersion in the triangular regions 
  above and below the disc, while $\sigma_{\mathrm{hor}}$ indicates the mean velocity
  dispersion in the disc plane.}
  \label{fig:vdispmap}
\end{figure}

\begin{figure}
  \centering
 \includegraphics[width=0.5\textwidth]{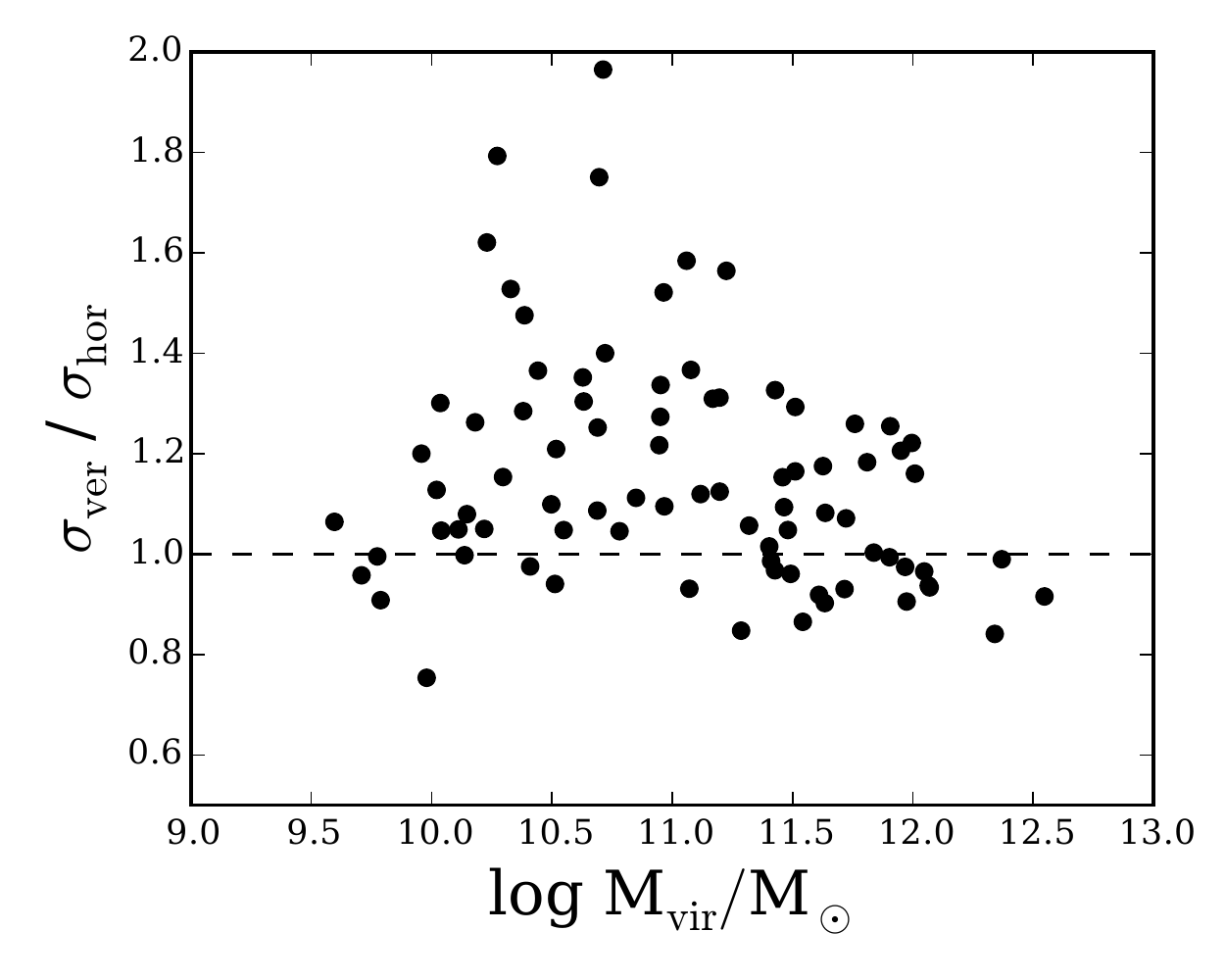}
  \caption{$\langle\sigma_{\rm ver}\rangle/\langle\sigma_{\rm hor}\rangle$ 
  as a function of halo mass, \mvir.  
  $\langle\sigma_{\rm ver}\rangle > \langle\sigma_{\rm hor}\rangle$ in $70\%$ of NIHAO galaxies, 
  which can be interpreted as a preference for outflows to be perpendicular to the disc. 
  Low and high mass galaxies have less clear bipolar outflows than galaxies with
  masses in the range $10.3 <\log(\mvir/$M$_{\odot})<11.3$. 
  Low feedback efficiency in low mass galaxies and deep potential wells in 
  high mass galaxies limit the amount of gas that escapes perpendicular to the disc.}
  \label{fig:sigmaratio}
\end{figure}

\subsection{Outflows and shape of the CGM}

\begin{figure}
  \centering
 \includegraphics[width=0.45\textwidth]{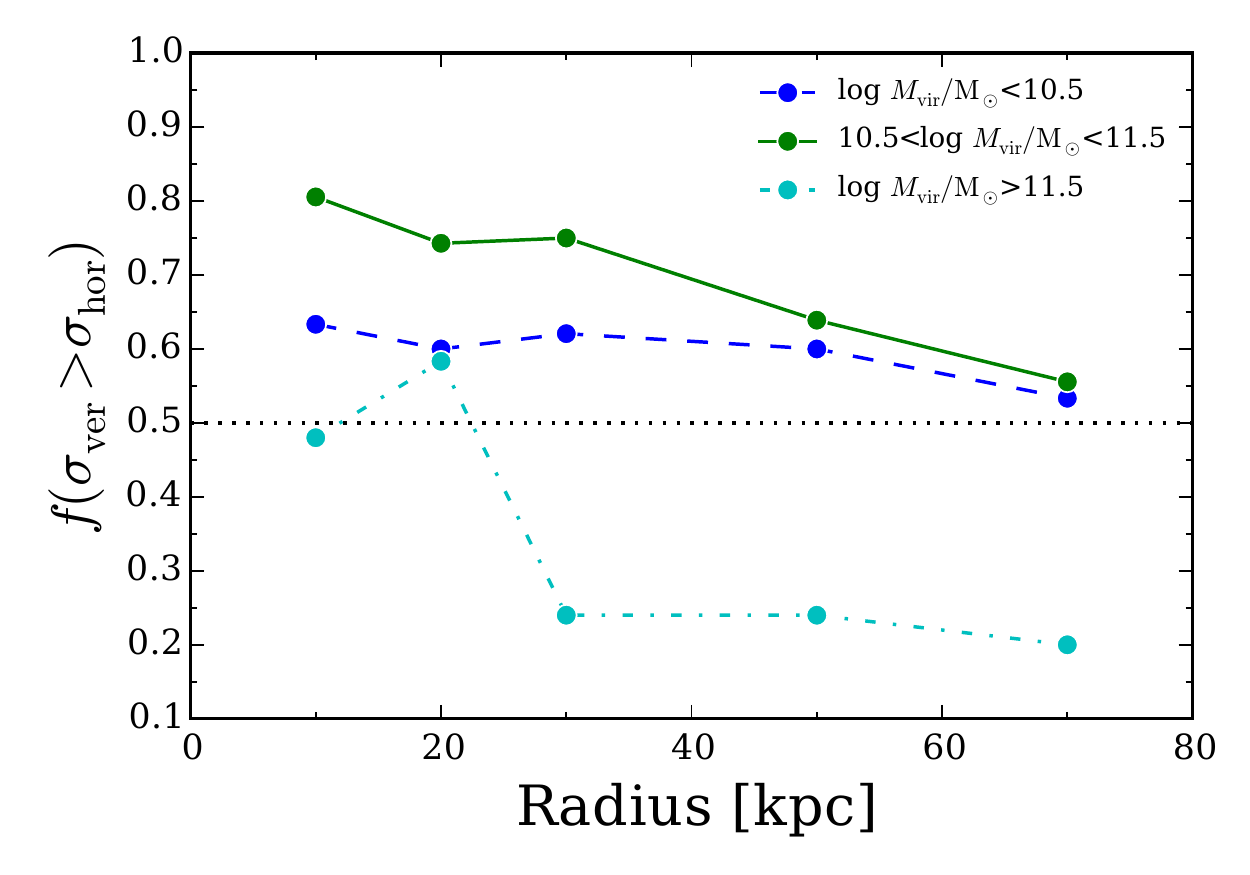}
 \caption{Fraction of all NIHAO galaxies with 
 $\langle\sigma_{\rm ver}\rangle > \langle\sigma_{\rm hor}\rangle$ 
 as a function of galaxy radius in three different halo mass ranges,
 $log \Mvir/\Msun <10.5$ (blue dashed line), $10.5<log \Mvir/\Msun
 <11.5$ (green solid line) and $log \Mvir/\Msun >11.5$ (cyan
 dash-dotted line). 
 The higher fraction of galaxies with 
 $\langle\sigma_{\rm ver}\rangle > \langle\sigma_{\rm hor}\rangle$ at small radii indicates that
  bipolar outflows extend to $\sim$40 kpc. The low fractions for the
  largest halos are due to the extended gas discs that dominate the
  velocity dispersion in the plane of the discs.}
  \label{fig:frac}
\end{figure}

Since quasars are so rare, it will never be possible to observe 
the shape of the low density CGM with absorption line studies.
However, closer to the disc, integral field unit (IFU) spectrographs
are making it possible to map velocity dispersions of gas immediately surrounding galaxies.
In recent observational studies \citep{Wild2014}, it has been assumed that 
an increase in velocity dispersion above and below the plane of the
disc is a dynamical signature of bipolar outflows. {Since outflows
  are expected to create steamers extending radially outward from the
  nucleus, the line-of-sight velocity dispersion should increase in
  the presence of a superwind. This is due to the increased line width
caused by the bulk motion of the gas toward and away from the
observer \citep{Wild2014}. \citet{Wild2014} use this method to study
the kinematical structure of the ``Mice'', a major merger galaxy pair,
and identified strong, galactic scale outflows in one the of ``Mice''
(NGC 4676A)
using the IFU survey CALIFA.}
Since a pressure driven wind will exit a galaxy in the direction of least pressure \citep{Hayward2015},
we expect outflows to exit our discs in bipolar outflows.

For this study, we make mock IFU observations of the disc edge-on. 
To this end, we project the galaxy onto the x-y plane and divide the plane into cells of approximately
the same number of particles. A Voronoi tessellation is calculated
based on the center of mass of each cell and the line-of-sight velocity dispersion is calculated for all particles
in each cell. 
An example of a velocity dispersion map for the inner 10 kpc of a $5\times10^{11} M_{\odot}$
galaxy from NIHAO is shown in Fig. \ref{fig:vdispmap}. 
The black dashed lines split the map into four triangular areas. 
The mean velocity dispersion of each of these is calculated.
The mean velocity dispersion about the center of mass of the cells in 
the vertical direction (above and below the disc) is called $\sigma_{\mathrm{ver}}$. 
The mean velocity dispersion in the cells along the plane of the disc (left and right) regions 
is called $\sigma_{\mathrm{hor}}$.

To understand whether there is an outflow trend with halo mass, Fig. \ref{fig:sigmaratio} 
shows the ratio of the mean velocity dispersions in the two directions, 
$\sigma_{\mathrm{ver}}/\sigma_{\mathrm{hor}}$, inside 10 kpc as a function of \Mvir.
The most significant outflows (i.e. the departure from a ratio of 1) are seen between $10.2<\log(\mvir\,/ {\rm M}_\odot)<11$.
These are similar masses to where the maximum stellar feedback effects are seen
in other studies like \citet{DiCintio2014} and \citet{Tollet2016}, 
so it is no surprise that the strongest bipolar outflow signal is seen here.

In the lowest mass galaxies, the lack of bipolar signal can be attributed to the lack of star formation. 
Stars form sporadically at these low masses, so 
it is unlikely outflows would be detected at the one output when these measurements are made. 
Few stars have formed over the last several Gyr in these lowest mass simulations.
In higher mass galaxies, there is plenty of star formation to potentially drive winds, 
but wind strengths do not increase so the deepening
potential well in more massive galaxies keeps the gas locked inside the disc.

A perfectly isotropic velocity dispersion would not show any of the heightened
velocity dispersion in the vertical direction that is seen in Fig. \ref{fig:sigmaratio}. 
An isotropic distribution would have half the galaxies above and half below 1.
Inside 10 kpc, $71\%$ of galaxies have an increased velocity dispersion 
above and below the plane of the disc relative to around the disc. 
So, Fig. \ref{fig:frac} looks at the fraction as a function of radius
{for three different halo mass ranges}. 
{Both the small (log $\Mhalo/\Msun<10.5$) and intermediate
  ($10.5<{\rm log} \Mhalo/\Msun<11.5$)
  halo mass ranges }shows that the velocity dispersion becomes more isotropic as we move out. 
Bipolar outflows are most visible in intermediate mass galaxies at radii
out to around 50 kpc. 
Beyond 50\,kpc, CGM turbulence and galaxy mergers mix the gas such that the 
bipolarity is no longer visible. {The largest halos (log $\Mhalo/\Msun>11.5$)
  have a very low dispersion ratio. This is due to the fact that our
  largest systems are dominated by extended gas discs, which fall into
our CGM definition but lie in the plane of the disc.

In the center of the halo ($r\lesssim50$\,kpc), our predictions of
the velocity dispersion will depend on our specific thermal feedback
prescription, since this is what is driving the outflows. And yet, the
feedback strength was calibrated to obtain the correct stellar mass -
halo mass relation, so although stronger or weaker feedback would
effect this result, such a prescription would not at the same time
also match the stellar mass relation.}

\begin{figure}
  \centering
 \includegraphics[width=0.49\textwidth]{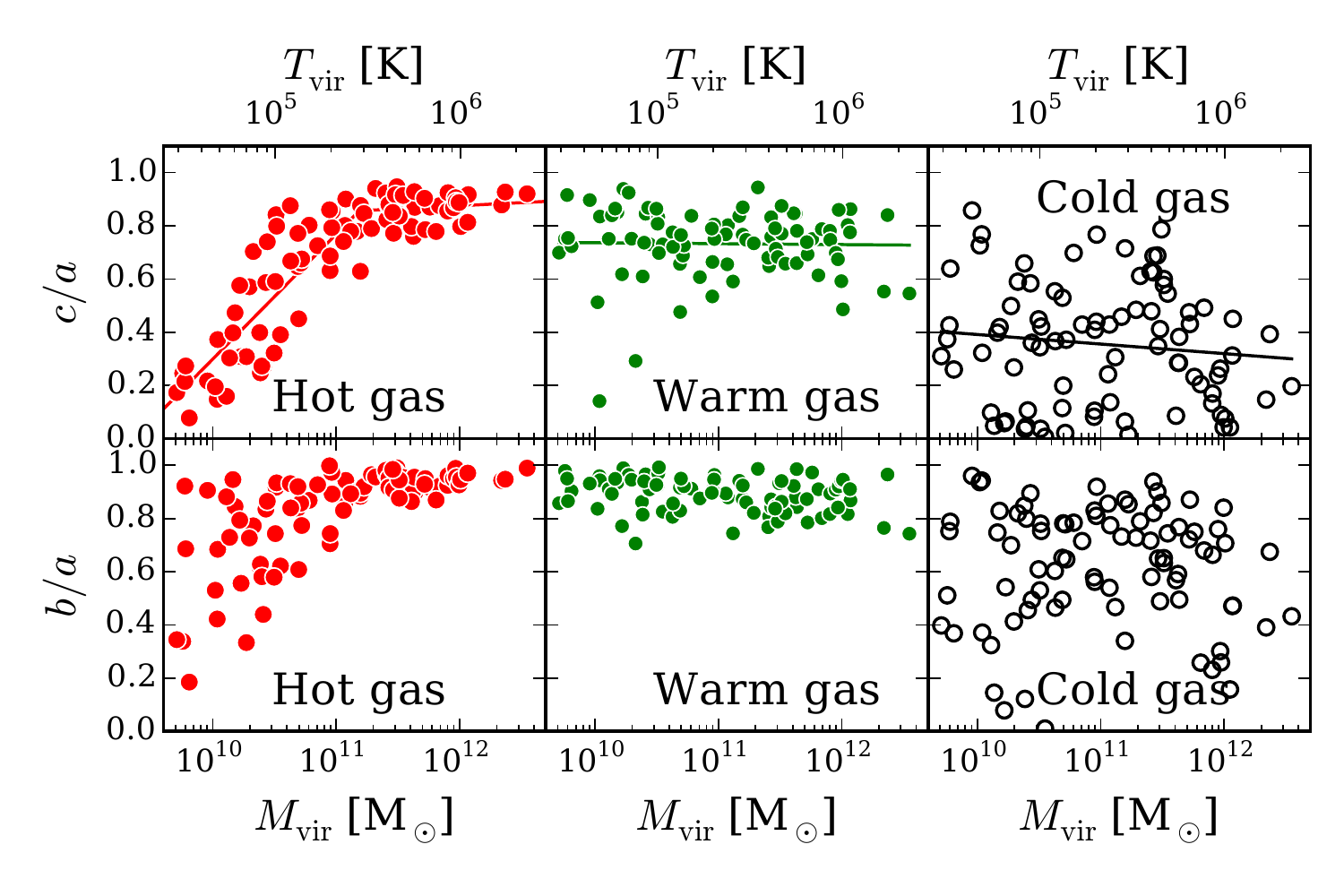}
  \caption{ $c/a$ (upper panels) and $b/a$ (lower panels)
    for the gas shape (without the disc in 0.5\rvir, this radius is a
    compromise between being too close, which would mean not capturing enough of the
  halo and too far, where the CGM is already so well mixed, that no
  structure would be recognisable.) in three temperature
    ranges. Hot gas: $T/K>10^6$ (red circles), warm gas: $3\times10^4<T/{\rm K}<3\times10^5$ (green circles) and
    cold gas:  $T/K<15000$ (black open circles). Lines show fits to the
    data in the respective color. 
    Fit values can be seen in table \ref{tab:cas}.}
  \label{fig:cas}
\end{figure}

\begin{table*}
\caption{Fitting parameters for the \mvir-$c/a$ relation in
  Fig. \ref{fig:cas}. Parameters are of the form $c/a = \alpha
  \cdot \log_{10}(\mvir) + $c.}
\begin{tabular}[c]{l|c|c|c}
\hline
\hline
& T/K & $\alpha$ & c\\
\hline 
Hot gas ($\mvir<2\times10^{11} M_{\odot}$) & $T>10^6$ & 0.469 & -4.395\\
Hot gas ($\mvir>2\times10^{11} M_{\odot}$) & $T>10^6$ & 0.023 & 0.596\\
Warm gas & $3\times10^4<T<3\times10^5$ & 0.004 & 0.771\\
Cold gas & $T<15000$ & -0.036 & 0.751\\
\hline
\end{tabular}
\label{tab:cas}
\end{table*}

\begin{figure}
  \centering
 \includegraphics[width=0.45\textwidth]{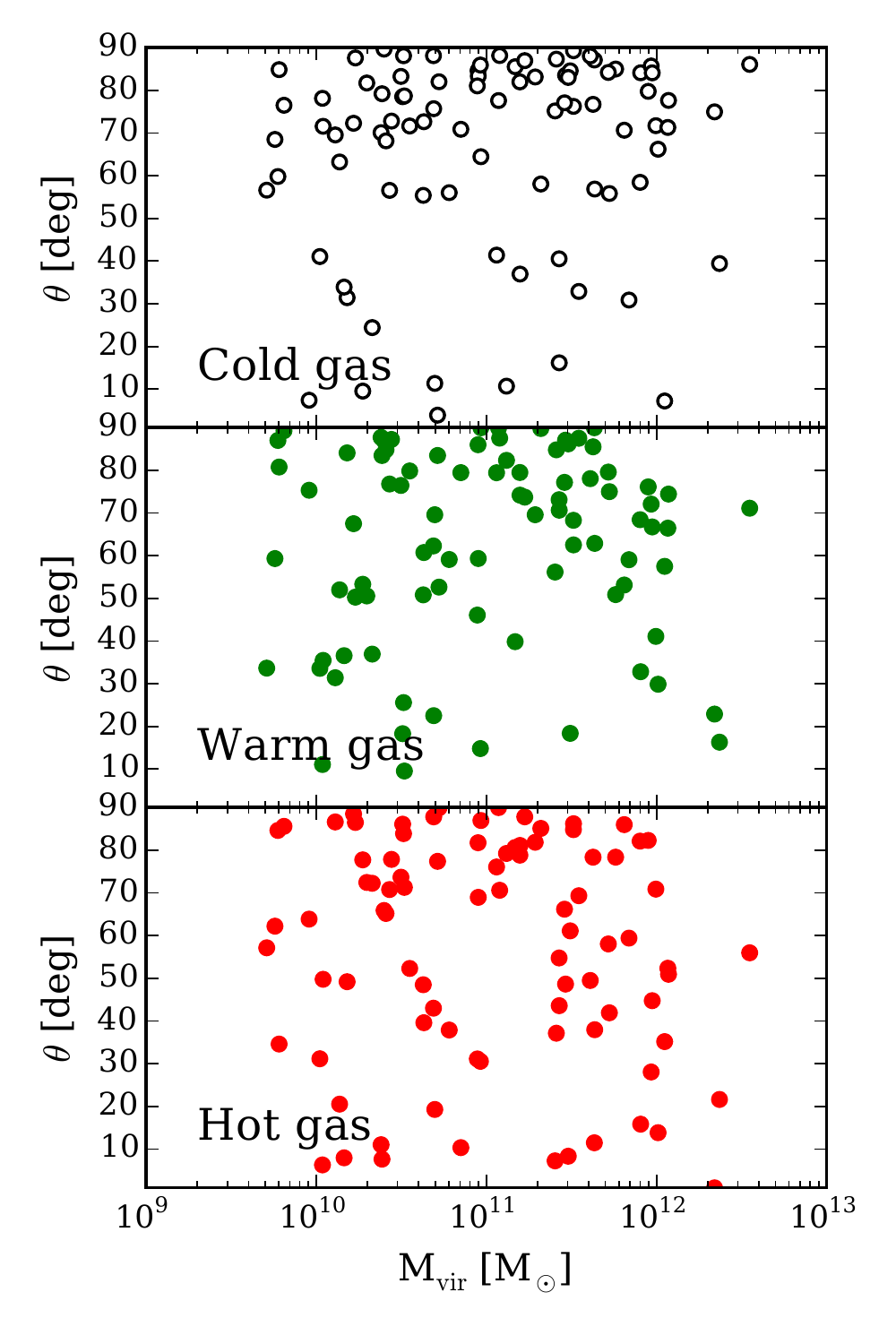}
  \caption{Angle $\theta$  between the rotation axis of the disc and
    the major axis of the inertia tensor fit as a function of viral
    mass. 90\textdegree~indicates that the major extent of the fitted gas
    is in the plane of the disc, while 0\textdegree~indicates gas
    extending perpendicular to the disc. Each panel shows a different
    temperature range of the gas, as denoted on the plots.}
  \label{fig:angles}
\end{figure}

\subsubsection{Triaxial fit}

IFU observations are limited to the inner few kpc of a galaxy
and QSO absorption spectra are limited to viewing one line of
sight. In light of this, it is very difficult to get any data on the
general shape of the CGM out to the virial radius. 
To get a theoretical handle on its shape {and provide some
  numerical estimates for further use by models}, we use the inertia tensor to give a rough quantity of the shape.
The inertia tensor is a $3\times3$ matrix, which in the common definition
 is the sum of tensor products of all the particle positions weighted by their masses.
The eigenvectors of the inertia tensor point in the 3 orthogonal directions of the
major, intermediate, and minor axes.
The eigenvalues of the inertia tensor are the lengths of the major axis and two orthogonal minor axes.  
These eigenvalues thus define a triaxial ellipsoid.

We wish to focus our study of CGM shape on only halo gas, so we try to exclude
disc gas from our shape measurement.  
Low density is the best determinant of halo gas, so we classify the halo 
as all gas with gas density, $n<10^{-2}$ cm$^{-2}$, and lies outside
0.1\rvir. {The density was chosen as the density where self-shielding starts becoming effective. We did not add a temperature cut because some analyses of the CGM have found that it is surprisingly cold. We did not want a temperature cut in our definition to preclude this possibility in our analysis.}

Fig. \ref{fig:cas} shows the eigenvalues of the inertia tensor for
halo inside 0.5\rvir.
The upper panels show  $c/a$, where a is the
major axis and c the minor axis of the fitted ellipsoid as a function of total mass of the galaxy. 
The lower panels show $b/a$, where b is the middle axis of the fit. 
The left panels show the hot gas (red circles, $T>10^6$K) in each galaxy. 
The right panels show the intermediate (green circles, $3\times10^4<T/{\rm K}<3\times10^5$ ) 
and cool gas (white circles, $T<1.5\times10^4$ K) for each NIHAO
galaxy. {The three temperature regimes are roughly the portions of the cooling curve dominated by 3 different cooling processes.
The hot regime is dominated by bremsstrahlung, the warm is dominated
by metal cooling, while the cool regime is dominated by hydrogen and helium line cooling.}
The lines represent simple fits to the distribution of $c/a$.
The fitting parameters are given in table \ref{tab:cas}.

The $c/a$ distribution of the hot gas rises steadily with mass up to
around $2\times10^{11} M_{\odot}$. We present the power law fits to the
$c/a$ ratios in \ref{tab:cas} of the appendix. Since the $b/a$ ratio also grows
with a similar, albeit not so strong trend, the hot gas distribution transforms from being prolate at
low masses (small $c/a$ and small $b/a$) to being mostly oblate or even spherical at high masses. 
 The flattening at high masses can be understood with galaxy
virial temperatures in mind. The virial temperature is the peak
temperature inflowing gas can reach falling into a potential well. 
Anything higher, so for most NIHAO galaxies our hot CGM gas, should come
from gas heated by stellar feedback. 

For the most massive galaxies, our hot CGM threshold is their virial
temperature. We expect more massive galaxies where the gas is heated by gravitational infall should be more spherical.
The hot gas already attains a spherical shape in galaxies with lower virial temperature than $10^6$ K.
The hot CGM may become spherical at slight lower masses (lower virial
temperatures) due to winds reaching
sufficient distances from the galaxy, and then being blown back and
shaped by the ambient CGM.

Warm gas does not exhibit the same trend with $c/a$ as the hot gas at low masses. 
Both the $c/a$ and the $b/a$ values stay relatively constant and high ($\sim0.8$) 
across all halo masses indicating a mostly spherical shape. 
The relatively high $c/a$ and $b/a$ values are due to the fact that this gas 
is both cooling out from the halo and isotropically falling inwards. 
Some of this is also gas that has been heated by stellar
feedback and blown out perpendicular to the plane of the disc. 

The cold gas exhibits the lowest $c/a$ values, as one would expect since 
it is dominated by the inner disc.  
One would further expect a cold gas disc to have high $b/a$ values since
those axes should be approximately the same length.  
For the most part, the $b/a$ values are high in Fig. \ref{fig:cas}, though
they tend towards lower values at low and high masses.

\subsubsection{CGM-Disc orientation}

Since outflows are expected to be aligned perpendicular to the plane of the disc, 
the hot CGM should appear bipolar in galaxies with lower virial
temperatures than $10^6$ K. The prolate shape at low masses could indicate that the hot CGM 
is configured as a bipolar outflow.
To understand whether this is the case, we present
Fig. \ref{fig:angles}. It shows the angle between the angular momentum vector of the disc and
the major axis of the inertia tensor.
The angle is plotted as a function of \Mvir. 
A 90\textdegree~angle indicates that the major moment of inertia axis lies in the plane of the disc, 
while 0\textdegree~indicates the gas is perpendicular to the disc.

There is no trend with mass for either the hot or warm gas.
In the angle analysis, the alignment appears random. Since the warm
and also the hot gas is for the most part spherically shaped, it is
not surprising that the orientation of the major axis of the inertia
tensor is random.

At low mass, some of the random alignment may not come from the outflow 
but from the disc angular momentum vector.
Stellar feedback drives so much turbulence that 
most NIHAO galaxies with $\mvir<10^{11} \msun$ do not have well 
ordered thin discs \citet{Obreja2015}, in effect making the direction given to the disc random.

The only temperature gas that does not have random scatter of angle values is the cold gas.  
For the cold gas, the angle clusters around 90\textdegree. 
Such a configuration is consistent with an extended cool gas disc.

The lack of a clear bipolar shape out to large radii is consistent with
our velocity dispersion analysis which showed a fading away of the
bipolarity at around 40 kpc. Even the smallest NIHAO galaxies have
virial radii larger than this.

\begin{figure}
  \centering
 \includegraphics[width=0.5\textwidth]{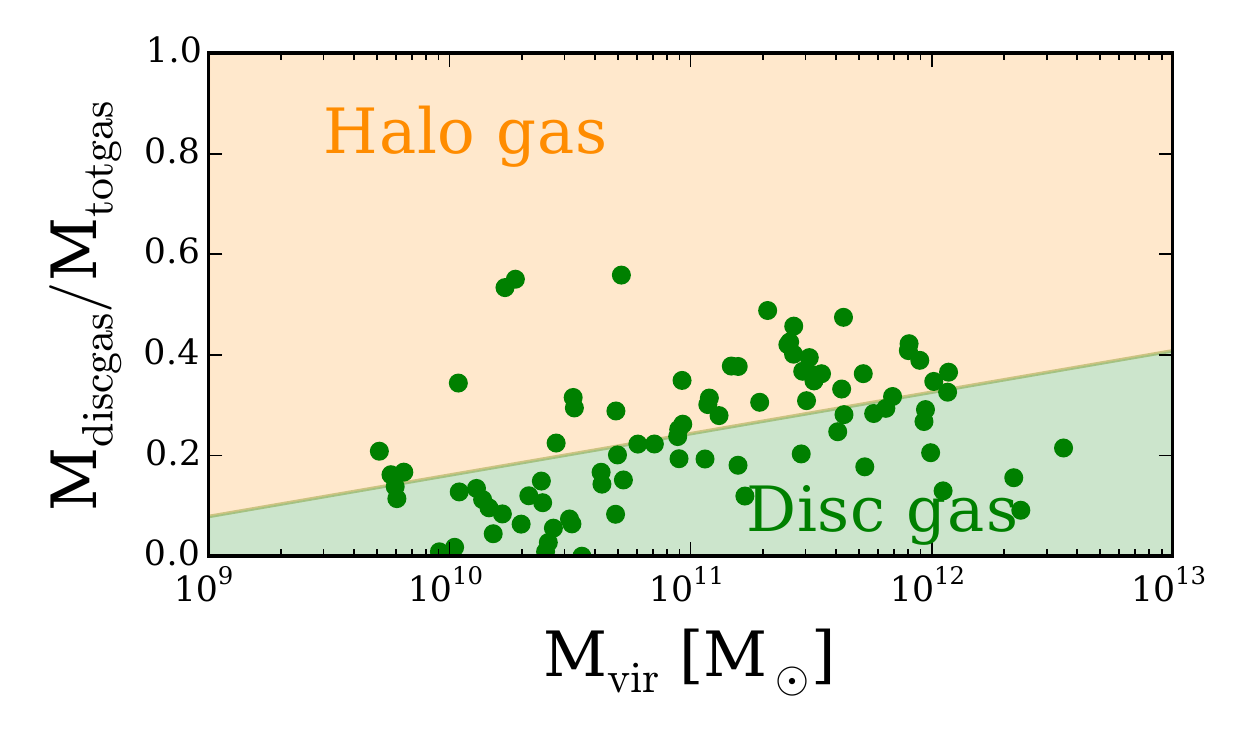}
  \caption{Mass fraction of disc gas relative to the total amount of
    gas. A power law fit to the data is shown as the boundary of the
    coloured areas. The parameters of the fit are given in table
    \ref{tab:gasfracs}.}
  \label{fig:discfracs}
\end{figure}

\begin{figure*}
  \centering
 \includegraphics[width=\textwidth]{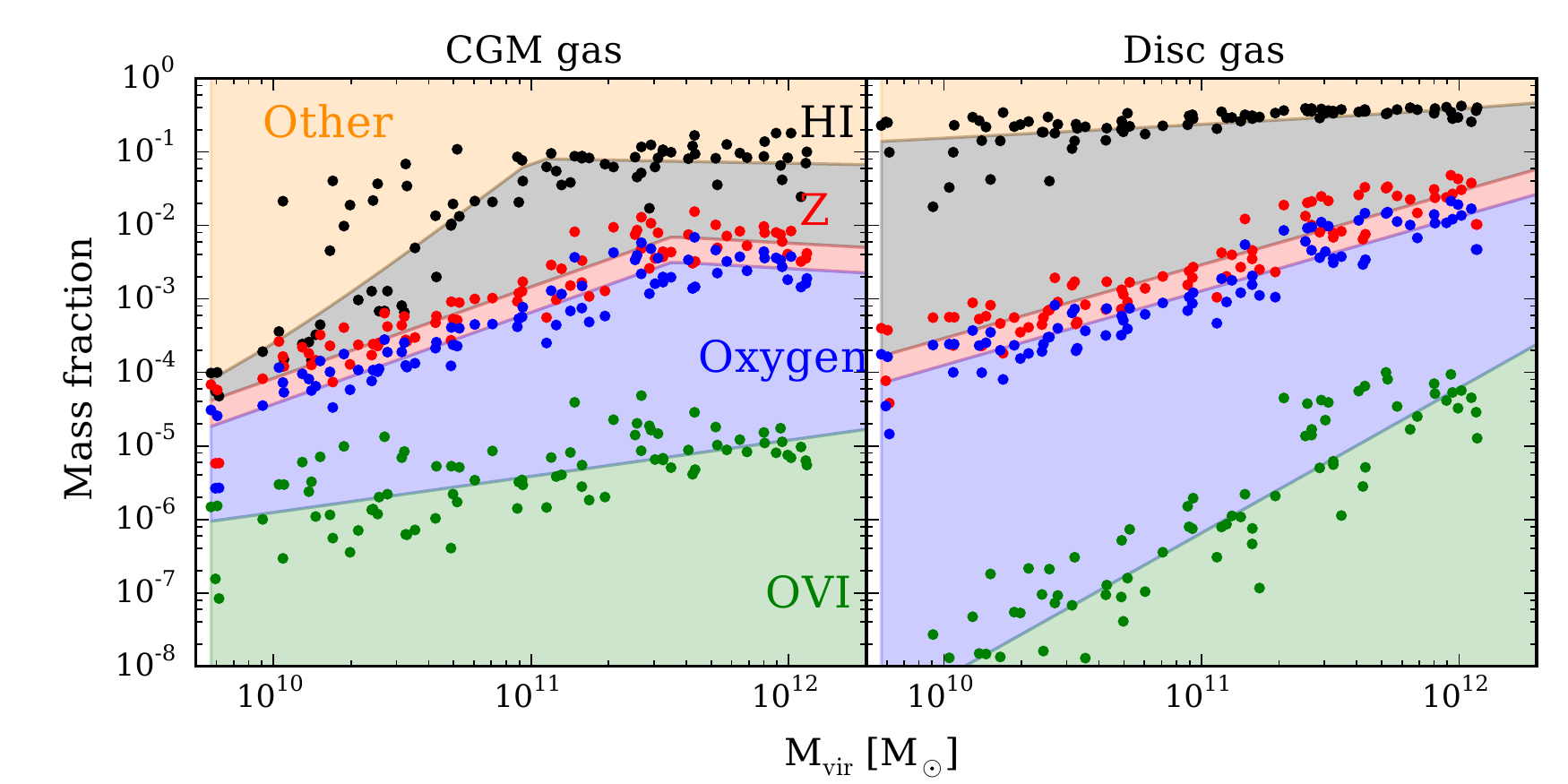}
  \caption{Mass fraction of gas split into chemical components. The
    plot should be read from top to bottom: Of all gas in a given halo
  mass range, the fraction of gas that is {\sc Hi} is given by the
  black line and the amount in metals is given by the red line. Of all
  metals, the mass fraction in oxygen is given by the blue line. Of
  all oxygen, the mass fraction in {\sc Ovi} is given by the red
  line. Each line is a fit to the data points and the fit parameters
  are given in table \ref{tab:gasfracs}.}
  \label{fig:gasfracs}
\end{figure*}

\subsection{Chemical composition of the CGM with galaxy mass}
\label{sec:chemcomp}

Lastly, we attempt to portray the chemical makeup of the {CGM} and the
disc separately and how these change as a function of halo mass in our simulations. 
We follow \citet{Suresh2015}  and define the disc as all gas inside 0.1\rvir\,
with a density higher than 0.01 cm$^{-3}$, and the {CGM} all other gas inside \rvir.  
Thus, {CGM} gas is characterised according to its diffuse nature, but the definition
allows satellite galaxies and streams that may have pockets of higher density
gas to be classified as part of the halo.
In this way, the definition includes all gas inside \rvir.

Fig. \ref{fig:discfracs} shows the disc gas mass fraction of all the gas inside \rvir.
The disc gas fractions are plotted as a function of \Mvir.
In nearly every case, the halo gas comprises the majority of the gas
inside \rvir\,(note the logarithmic axes),
and in some cases even more so.
The disc gas fraction shows a large amount of scatter. 
Below $3\times10^{10} \msun$, the disc gas fraction varies from almost nothing to over 50\%.
The high variation corresponds with the bursty star formation 
that happens in such low mass galaxies.  
The high disc gas fractions represent galaxies that haven't had much
recent star formation,
while the low disc gas fraction galaxies have had a starburst that
blew their cool, dense disc gas away.
The NIHAO sample includes few galaxies over $M>10^{12} \msun$,
but the three that show up in Fig. \ref{fig:discfracs} indicate that the disc
gas fraction starts to drop above $10^{12} \msun$.
The pattern is reminiscent of galaxy formation efficiency found
using abundance matching techniques \citep{Guo2010,Moster2013,Behroozi2013}.
One expects the disc gas mass to appear similar to star formation efficiency
since stars must form out of cool, dense gas.

Fig. \ref{fig:gasfracs} shows the mass fractions of the cool and hot gas tracers,
\ovi\, and \hi, out of the total gas masses for the halo and disc gas components separately.
Since \ovi\, is a particular ionic species of oxygen, which itself is a metal, 
the colored \ovi\, and oxygen regions are subfractions of the larger region of which they are part.
The region of the plot above the ``Z'' metals line is comprised of hydrogen and helium.
The part labeled ``\hi'' is the neutral hydrogen component.
For clarity, we note that the \hi\, line and points are 
the \hi\, mass fraction added to the metals mass fraction (M$_{{\rm HI}}$ + M$_{Z}$)/M$_{\rm tot, X}$.

The behavior of the halo and disc gas fraction differ as a function of mass.
In the halo, the fractions of the species we tracked increase with increasing \mvir\,to
some threshold mass at which point they flatten out and stop rising as a function of mass
and may start to decline slightly. 
In the disc, the fractions of each component continues rising steadily through
the high end of the NIHAO mass range.

The metals are all produced as the result of star formation, 
so one expects the metal fraction to increase in galaxies forming stars.
Except for the extreme low mass end, the NIHAO galaxies form stars throughout their lifetimes.
The continued increase in the disc metal gas fraction reflects the continued star formation.
{The halo metals gas fraction flattens because metal enriched outflows decrease significantly, 
or stop altogether, once galaxies surpass some mass threshold as it becomes harder for winds to escape the disc in
the more massive gravitational potential wells \citep[see][for a detailed study of this]{Muratov2015}.} { Based on this, the
threshold halo mass at which the flattening occurs is a tug-of-war
point between feedback strength, which pushes gas outward, and the
depth of the gravitational well. As such, if this turnover mass were measured
observationally, it could set a contraint for the strength of feedback models.}

The metals reach the mass threshold where 
their mass fractions flatten around $3\times10^{11}\msun$.
\hi\, flattens at a lower mass around $\Mvir=10^{11}\msun$.
Different physics governs the \hi\, fraction, since it does not increase
due to of star formation.
Its fraction is dominated by ionization; ``other'' gas in the halo is mostly ionized hydrogen.
Whether the \hi\, fraction comes from outflows or is simply a natural consequence of
the gas density profiles of galaxies is a question for study in future work.
The sharp decrease in \hi\, fraction in some of the lower mass galaxies reflects the lack
of any dense gas in the halos of those galaxies.

The \ovi\, trend is ambiguous as it has the largest scatter of any species,
so if the high boundary is indicative of the \ovi\, fraction, then the \ovi\, fractions reduces.
The \ovi\, can also be well fit using a single power law, which we have done for simplicity.
\ovi\, is a similar case to \hi\, since it will depend on the ionization conditions.  
The virial temperature of the most massive halos is around the
ionisation temperature for \ovi. 

While it is hard to see on the logarithmic scale, 
the disc \hi\, fraction grows by a factor of a few over the halo mass range. 
That trend follows the the disc gas mass fraction, 
and reflects the increasing gas densities with \Mvir.

The \ovi\, fraction in the disc increases the most steeply of any species, with a slope 
$M_{\rm disc,OVI} \propto \Mvir^2$.  
That increase reflects the deeper disc potential, which is able to hold onto more hot gas as
the galaxy gets more massive.

To each mass fraction, we fit power laws. 
In the case of the gas in the halo, we fit a broken power law to the \hi, the metallicity and
the oxygen components. 
The threshold for the break was chosen ``by eye'' and is $10^{11} \msun$ for \hi\, 
and $5\times 10^{11} \msun$ for both the metallicity and the oxygen components. 
The fitting parameters are given in table \ref{tab:gasfracs}.

\begin{table}
\caption{Fitting parameters for the power law fits in
  fig. \ref{fig:gasfracs}. Parameters are of the form
  $\log_{10}(f_{\rm X}) = \alpha
  \cdot \log_{10}(\mvir) + $c.}
\begin{tabular}[c]{l|l|c|c|c}
\hline
\hline
&Name & \mvir\,range & $\alpha$ & c\\
\hline 
Halo gas &{\sc Hi} & $<1\times 10^{11} \msun$ & 2.643 & -30.202 \\
&{\sc Hi} & $>1\times 10^{11} \msun$ & -0.050 &  -0.589\\
&Z & $<5\times 10^{11} \msun$ & 1.214 & -16.208\\
&Z & $>5\times 10^{11} \msun$ & -0.309 & 1.476\\
&O & $<5\times 10^{11} \msun$ & 1.219 & -16.608\\
&O & $>5\times 10^{11} \msun$ & -0.315 & 1.202\\
&{\sc Ovi} & all & 0.492 &  -10.823\\
Disc gas &{\sc Hi} & all & 0.183 &-2.643\\
& Z & all & 1.00 & -13.534\\
& O & all & 1.011  & -14.013\\
&{\sc Ovi}  & all & 1.972 & -27.880\\
Disc/Total & & all & 0.082 & -0.662 \\
\hline
\end{tabular}
\label{tab:gasfracs}
\end{table}

\section{Discussion}
\label{sec:discussion} 
In general, the simulated CGM compares well with current observations,
but two problems stand out: 
1) the \ovi\, simulated column densities are almost 1 dex lower than observations;
and 2) optically thick \hi\, extends much further in observations than simulations.

\subsection{\ovi\, deficit}

{We use the suite of 86 NIHAO simulations to analyse
the CGM and the extent of \ovi\, column densities out past the virial
radius of galaxies from log(\Mhalo/\Msun)$\sim 9-12$.
The right panel of Fig. \ref{fig:mstar_ovi} shows how the predicted \ovi\,
column density changes with the stellar mass of the
galaxy. Fig. \ref{fig:lstar_ovi} shows the changes with impact
parameter and luminosity
and includes the observations. We underpredict \ovi\, at all radii and
for all galaxy luminosities by about the same amount everywhere,
approximately 1 dex. The origin of this discrepancy is not yet well understood, but has been seen in most
galaxy formation simulations to date
\citep{Hummels2013,Suresh2015,Ford2015}. 
While \citet{Suresh2015} suggest that the observations are consistent
with a constant column density across galaxy mass/luminosity, {the
observations are also consistent with} an increase to around $1L^*$, above which the
trend flattens off. }
In the simulations, the plateau is at log$(N_{\rm OVI}/{\rm cm}^{-2})\sim13.5$ in each radial range.
The plateau is reached at progressively higher luminosities at larger radii.

The observational data is relatively sparse and shows a fairly broad scatter, but
there may be a similar plateau that is higher than the simulations, at log$(N_{\rm OVI}/{\rm cm}^{-2})\sim14$.

{\ovi\, can be photoionised at low temperatures and low densities
($T\lesssim10^{4.5}$\,K, $\rho\lesssim10^{-4}{\rm cm}^{-3}$) by incoming radiation assuming
photoionisation equilibrium  \citep[see][for possible shortcomings of
this assumption]{Kollmeier2014} and a constant UV
background. Assuming collisional ionisation equilibrium produces \ovi\, in denser
gas ($\rho\gtrsim10^{-4}{\rm cm}^{-3}$) with temperatures around
$10^5-10^6$K. Disentangling these two channels is vital to get the correct amount of \ovi.}

\begin{figure}
  \centering
 \includegraphics[width=0.5\textwidth]{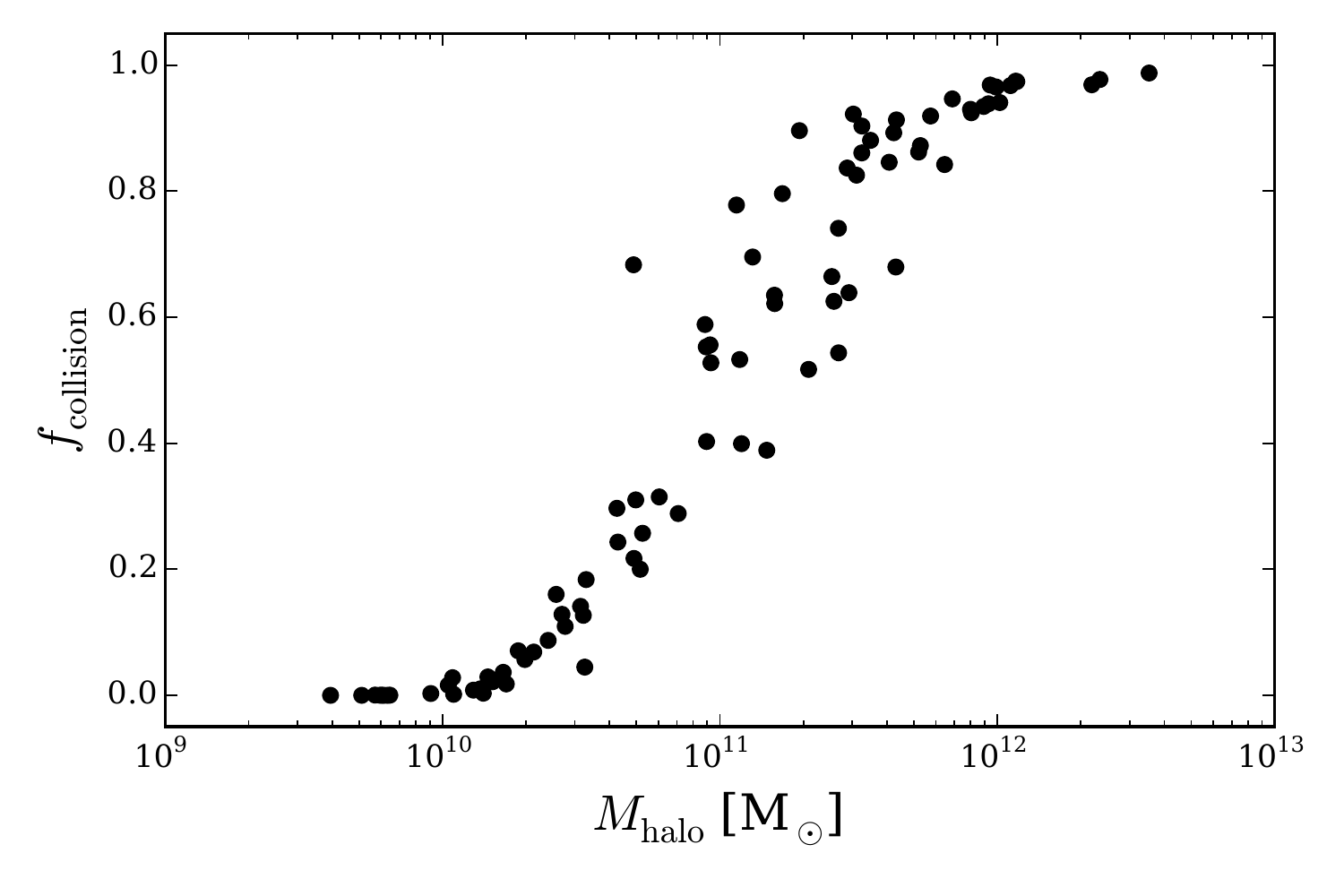}
  \caption{Mass fraction of \ovi\, in the CGM created by collisional ionisation per
    galaxy as a function of halo mass. \ovi\, in low mass galaxies is
    mainly created through photoionisation from the UV
    background. Massive galaxies have a large fraction of \ovi\, that is
  collisionally ionised.}
  \label{fig:photo_coll}
\end{figure}

{Fig. \ref{fig:photo_coll} shows the mass fraction of \ovi\, in the CGM
produced
by collisional ionisation, $f_{\rm collision}$, as a function of halo
mass for all NIHAO galaxies. The definition of CGM is the same as in
the rest of the paper and $f_{\rm collision}$ is defined as the
fraction of \ovi\, above $10^5$\,K with respect to the total amount of \ovi\,
in the CGM per galaxy. This approximation of the amount of
collisionally ionised \ovi\, is adequate to get a sense of the trend
as a function of halo mass, which is all we are interested in
here. 

Small galaxies are mainly photoionised, while massive
galaxies are predominantly collisionally ionised. The approximate
threshold halo mass between these two regimes is $10^{11}$ \msun. This
is caused by the fact that much of the CGM gas is around the virial
temperature, which enters the \ovi\, collisional ionisation regime at
about this halo mass, being too low for smaller galaxies. 

The assumption of collisional ionisation equilibrium (CIE) generally works best in high density, low temperature regions. Not
assuming CIE has been tested in a variety of non-equilibrium models
\citep[see][]{CenFang2006, Savage2014, Oppenheimer2016}.
\citet{CenFang2006}  show a decrease in the \ovi\,
column densities relative to CIE since the shock heated gas is
initially above $10^6$\,K and cools
through the collisional ionisation
regime. Conversely, \citep{Savage2014} show that non-CIE
models can increase the amount of \ovi\, produced through collisional
ionisation at temperatures below $10^5$\,K. Although testing non-CIE
models might hold the answer to the \ovi\, deficit, it is
beyond the scope of this paper.

}

\citet{Ford2015} point out that while the virial temperature of massive halos around 
$10^{12} \msun$ is around $10^6$K, 
most gas that is {\it photo}ionized into the \ovi\ state has a temperature around $10^{4.5}$K. 
\ovi\, ionisation might best be explained as a transient state of hot gas cooling out of the halo. 
Simulations have a hard time capturing non-equilibrium states, 
so they might therefore show a persistent deficit in \ovi. 
\citet{Rahmati2015} find that collisional ionisation is a non-negligible effect for \ovi.  
Collisional ionisation fractions are above 20\% for $z=0.1$ in their
fiducial model.

{ The only way
photoionisation from the UV background could be the dominant process
in
producing \ovi\, {in massive halos above $10^{11}$\,\msun} would be if the CGM gas were systematically
cooler. This means either the gas never shock heated to the virial
temperature or the cooling time is short enough to pass through the
collisional ionisation range without exciting \ovi. But since
simulations generally have trouble with the overproduction of stars, a
further enhanced cooling time would worsen this problem.

Both \citet{Shen2012} and \citet{Suresh2015} ran models that included
local sources of photoionisation to test whether this significantly effects the overall amount of \ovi. Both found that the effect is negligible for
$r\gtrsim 50$\,kpc. So, although this could ameliorate the \ovi\,
deficit in the center of haloes, this is not sufficient to account of
the large amounts at larger radii. }

The two main pieces of physics that contribute to the \ovi\, column density are
{\it stellar feedback} that drives galactic outflows and the {\it radiation field}.
The detailed interaction of stellar feedback with the interstellar medium and
the CGM are poorly resolved in simulations.  The shock fronts occur
on small scales relative to the distance that the outflowing gas travels, so they are difficult to model.
In spite of the poor resolution, S12 and \citet{Hummels2013} showed 
the strength of stellar feedback has a big impact on \ovi\, column
densities. {Also, if resolution was a major cause of error, the
  NIHAO simulations would show a clear difference in CGM properties
  with halo mass, since the physical resolution decreases with increasing
  galaxy mass. }

Fig. \ref{fig:gasfracs} shows that there is plenty of oxygen in the CGM,
so matching the observations seems not to be a matter of transport, but one of getting it into 
the correct ionization state. {\citet{Hummels2013} tested manually
varying either the density, the metallicity or the temperature of their
simulations. Only
the simulations with feedback and outflows and a manually decreased
temperature (to $3\times 10^5$\,K) could produce enough \ovi\, to
reproduce the observations.
They note that temperature structure of the CGM is dependent on
resolution within $r\gtrsim 50$\,kpc based on their thermal feedback prescription.}

{An intrinsic error in our method to be noted is that we only
  account for gas absorption within $\sim4$\,Mpc of the galaxies. We are
  aware that quasar sightlines pass through a
larger volume, possibly gathering additional absorption lines along
this path. This would cause our column densities to be underestimated. We assume this effect to be small, similar to other
analysis such as \citet{Hummels2013}.}

\subsection{Extended \hi}
\label{sec:coolgas}

One of the more mysterious discoveries about the CGM has been observations of
optically thick, cool, neutral gas covering large cross sections surrounding galaxies \citep{Hennawi2006,Prochaska2013,Werk2014}.
Particularly perplexing is that the cool gas covering fractions increase
when looking at more massive galaxies that are also quasars \citep{Prochaska2014}.
Naively, more massive galaxies should have hotter CGMs and quasars
should ionize any neutral gas.  
Observations show the opposite: 
quasars have more cool, neutral gas in the form of small, dense clouds.

Our simulations are not ideally suited to study the CGM of massive galaxies at high $z$;
NIHAO is focused on galaxies smaller than the Milky Way at $z=0$.
However, we present a comparison to give some idea of how our simulations
perform on this problem. {Our \hi\, ionisation fraction takes
  self-shielding into account, which has the effect of increasing the
  amount of \hi\, in high density regions.}
Fig. \ref{fig:covfrac} shows that optically thick \hi\, covers a small fraction of \rvir\,at $z=0$, but
that the NIHAO simulations cover a large fraction at higher redshift that approaches
the observed value.
While the fraction of \rvir\,covered increases with redshift, 
Fig. \ref{fig:covfrac} also shows that the physical area covered changes little.

Fig. \ref{fig:r_hi} shows that the simulations match the cool
gas column densities very well in the inner regions of the CGM.
The simulations have trouble producing high enough \hi\, column densities
in the CGM outskirts.

{This is a general discrepancy between observations and simulations
that has not been solved: why there is so much cool gas at such large
distances from galaxies?}
In an evolving galaxy there are two possibilities: cool gas clouds are ejecta from the disc
or gas cools at large distances and is in the early stages of accreting onto the galaxy.

Since the cool gas is not pressure supported, cooling condensations
contribute to overcooling in simulations as the clouds fall onto the disc. 
It remains a mystery how clouds like those observed continue to be
supported at large distances in the CGM, since they do not appear to contribute to star formation.

SPH simulations without a careful numerical treatment of
hydrodynamics will overestimate these cooling condensations \citep{Maller2004}.
Such condensations are the numerical embodiment of the \citet{Field1965} instability.
\citet{Agertz2007} identified such blobs as a numerical problem in SPH 
that does not exist in grid simulations, where cool clouds are quickly destroyed 
when moving through a hot medium.

Hints to solve the mystery are coming from focused, high resolution simulations.
\citet{Quataert2015} showed that magnetic fields can provide internal pressure for cool clouds
and allow those cool clouds to survive for much longer times.  
%
High resolution simulations are also being used to understand 
how cool gas can be driven out of galaxies.
\citet{Scannapieco2015} simulated cold gas clouds in supersonic flows and found that 
they survived longer than clouds in subsonic flows because 
the shock front suppresses the Kelvin-Helmholz instability.
Such enhancements allow cold clouds to survive out to several
kiloparsec from galaxies, 
but not hundreds of kiloparsec where \hi\, is observed.

\citet{Brueggen2016} adds thermal electrons to the \citet{Scannapieco2015} simulations.
Conduction efficiently heats the cloud below a threshold density. 
When the cloud is more dense, it collapses into a dense filament that is
able to radiate away any energy that conducts into its center, which also lengthens its lifetime.
%
Adding to prior research, \citet{Thompson2015} take into account the rapid destruction of 
cool clouds entrained in a hot $\sim10^7$ K outflow.
They suggest that the cool gas in the CGM outskirts has two sources.
An initial source is hot outflowing gas that radiatively cools.
A secondary source is the gas that cools out of the shock front created as 
the wind plows into the CGM.

These modelers suggest that galactic winds include physics that is too 
difficult for simulations to model and provide prescriptions to include the physics in simulations.
It is unclear whether it is possible to easily include such prescriptions.
\citet{FaucherGiguere2015} has shown promising results in 
fully self-consistent simulations using higher resolution.
For instance, higher resolution adds a significant area of optically
thick gas.  {\citet{Rahmati2015} were also able to match the \hi\, column
  densities even at large distances from the galaxy using strong
  stellar and AGN feedback.}
Combining insights from both high resolution models and more advanced simulations
can help to deepen our understanding of how the CGM is created and evolves.

\section{Summary}

We analysed the CGM of the NIHAO galaxy simulation suite, 
which includes 86 galaxies ranging in halo mass from $10^{9.7}-10^{12.5}$ M$_{\odot}$.
The galaxies follow the stellar mass--halo mass relation \citep{Wang2015}
and the disc gas--stellar mass relation \citep{Stinson2015}. 

The hot and cold gas phases were examined using the commonly observed \ovi\, and \hi\, ions.
We compared their column density profiles with observations, 
studied the covering fractions of dense \hi, looked at the shape of the CGM and 
studied its chemical composition. 
Our goal was to present the simulations so that observers and 
modellers can compare with them in future studies. 

We combined column densities of \hi\, and \ovi\, from all of the galaxies in NIHAO
according to both the luminosity of the galaxy and impact parameter of the measured
column density. 
Overall, the simulations provide a good match to observations.
The column density of \ovi\,  increases as a function of galaxy luminosity to a plateau.
Observations are still too sparse to make definite conclusions, but 
the simulations seem to follow the same functional form.
Using an updated self-shielding approximation \citep{Rahmati2013}, 
the simulated \hi\, column densities follow observations more closely than \ovi.
At $z=3$, optically thick \hi\, covers the same fraction of \rvir\,as the observations.

In addition to the successes of the simulations, two problems common to 
many numerical simulations are apparent. 
While the trend of the \ovi\, simulations with luminosity agrees with observations,
its normalization is about 0.7 dex too low, a trend commonly seen in 
simulations \citep[e.g.][]{Hummels2013,Ford2015,Suresh2015}.
In S12, this discrepancy was not as apparent, since
they examined just two galaxies and used a slightly stronger
feedback model than what was used in NIHAO.  
We do not propose using stronger feedback to create more \ovi\,
because it would reduce star formation too much.
Instead, we point to the large amount of oxygen present in the CGM
and conclude that our \ovi\, ionization fraction is
incorrect, possibly due to the temperature structure of the CGM.


Second, the extent of optically thick \hi\, (log $N_{\rm HI}>17.2$) 
covers too small a fraction of the projected area, particularly between \rvir\,and 2\rvir. 
The source of the large amount of cool gas observed in the outer CGM
and how it stays out there remains unknown.
Unfortunately, our simulations shed little new light on the mystery.
We simply note that that optically thick \hi\, covers the same physical area 
as a function of galaxy halo mass at all redshifts in the simulations.

Given the similarity of the simulations and observations, 
we extend our analysis to the outflows that transport gas and metal enrichment to the CGM.
Specifically, we recreate observations of gas velocity dispersion around edge-on discs.
\citet{Wild2014} proposed that heightened velocity dispersions perpendicular to the disc
with respect to the velocity dispersion in the disc plane are evidence for bipolar outflows.
We detect a signal in intermediate mass galaxies ($10\lta\log \Mhalo/\Msun\lta11.5$),
though the ratio never exceeds two. 
Such enhanced velocity dispersions are visible in maps out to around
40 kpc in this halo mass range. 
40 kpc is the same distance to which \citet{Bordoloi2014} observes outflows. 
Beyond 40 kpc, the velocity dispersion signal is isotropic. Larger halos do not show this signal,
since their inner CGM is dominated by an extended gas disc which lies
in the plane of the stellar disc.

Given the weak outflow signal, it is perhaps no surprise that it is hard to find
any evidence for bipolar shapes in our simulations.
Using the inertia tensor, we studied the shape distribution for three different gas temperatures.
The hot ($T>10^6$ K) gas showed a somewhat bipolar shape in the lowest mass simulations.
The direction of the long axis was uncorrelated with the disc angular momentum axis.
So, at $z=0$, bipolar outflows do not make up a dominant mass component of the CGM.
Rather, the CGM shape is mostly spherical because of virialization and 
the interaction of outflows with the ambient CGM. 
The cool ($T<10^4$ K) CGM is an extended disc, 
while warm gas ($3\times10^4<T/{\rm K}<3\times10^5$) is always spherical.

The chemical enrichment of both the disc and the halo 
follow expected monotonically increasing trends. 
The mass fractions of \hi, metals, oxygen and \ovi\, can be well fit 
with power laws as a function of halo mass.
The power law fits might be useful in galaxy formation models 
that do not use full hydrodynamics, such as semi-analytical models. 
Most notably, there is much more oxygen than what is shown as \ovi, so fixing the \ovi\,
problem is not a transport problem, but an ionisation problem.

Our analysis here focused on the $z=0$ universe. 
More work is necessary to put these results into context of the high redshift universe.

\section*{Acknowledgments} 
We thank the anonymous referee for a helpful report.
We also thank Aura Obreja, Tobias Buck, 
Jonathan Stern, Joe Hennawi, and Glenn van de Ven for
helpful suggestions and useful conversations.

The NIHAO simulations were run using the galaxy formation code
GASOLINE, developed and written by Tom Quinn and James Wadsley.
Without their contributions, this paper would have been impossible.

TAG, GSS and AVM acknowledge funding by Sonderforschungsbereich SFB
881 ``The Milky Way System'' (subproject A1) of the German Research
Foundation (DFG).

The analysis was performed using the pynbody package
\citep[http://pynbody.github.io/,][]{Pontzen2013}, written by Andrew Pontzen and Rok Ro\v{s}kar 
in addition to the authors. 

The simulations were performed on 
the {\sc theo} cluster of the Max-Planck-Institut f\"{u}r Astronomie, 
the {\sc hydra} cluster at the Rechenzentrum in Garching, and 
the {\sc Milky Way} supercomputer, funded by the Deutsche Forschungsgemeinschaft (DFG)
through the Collaborative Research Center (SFB 881) ``The Milky Way System'' (subproject Z2), 
hosted and co-funded by the J\"{u}lich Supercomputing Center (JSC).

\bibliographystyle{mn}
\bibliography{bib}
\balance

\bsp

\label{lastpage}

\newpage
\onecolumn
\appendix
\section{}
\label{sec:App}

\begin{figure*}
  \centering
 \includegraphics[width=\textwidth]{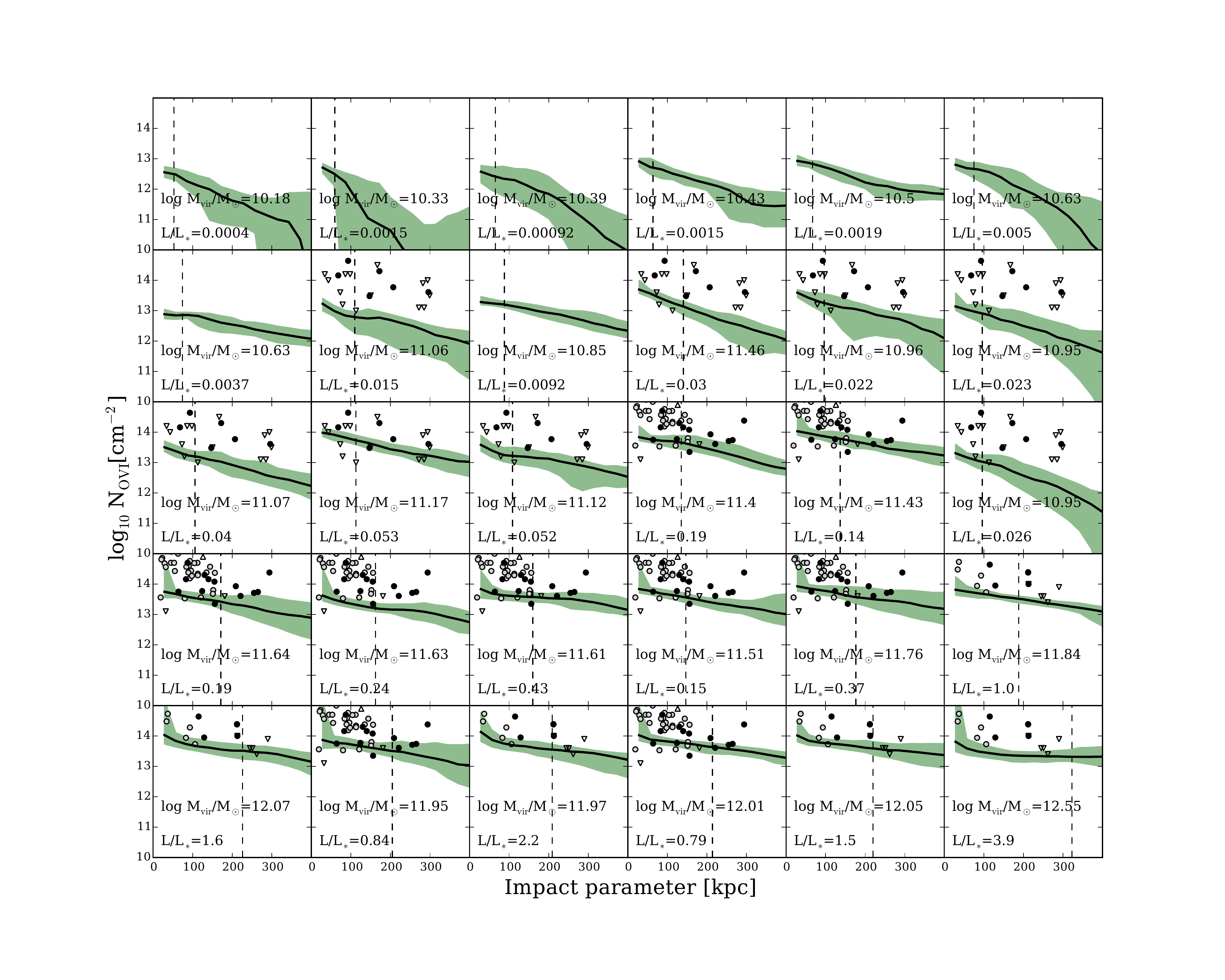}
  \caption{Mock sightline distributions of \ovi\, column density for
    the same 30 NIHAO galaxies as in fig. \ref{fig:ovimap} and fig. 3
    and 4 of \citet{Wang2015}. Black solid lines indicate the median column
    density for a given impact parameter bin. The green-gray shaded
    area represents the 5-95 percentile range. Observational data from Prochaska et al. 2011 (black symbols) and
    Tumlinson et al. 2011 (white circles). The Prochaska et al. 2011
    data is split
    into three groups: $0.01L*<L<0.1L*$, $0.1L*<L<1L*$ and $L>1L*$. Only the
    relevant group is overplotted according to the luminosity of the
    shown galaxy.}
  \label{fig:oviprof}
\end{figure*}

\begin{figure*}
  \centering
 \includegraphics[width=\textwidth]{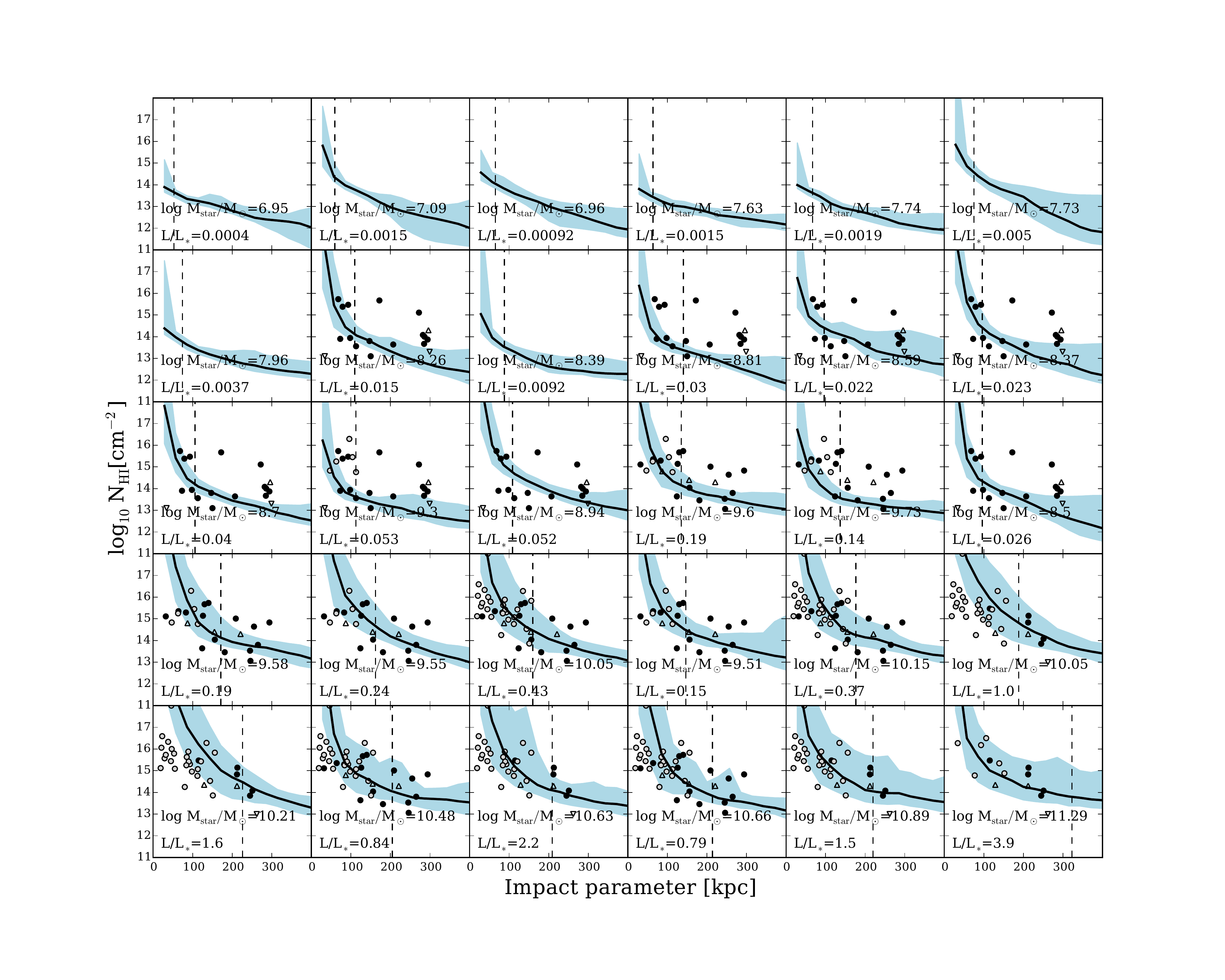}
  \caption{Mock sightline distributions of \hi\, column density for
   the same 30 NIHAO galaxies as in fig. \ref{fig:ovimap}. Observational data from Prochaska et al. 2011
    (black symbols) and
    Tumlinson et al. 2013 (white circles).  The
    data is split
    into three groups: $L<0.1L*$, $0.1L*<L<1L*$ and $L>1L*$. Only the
    relevant group is overplotted according to the luminosity of the
    shown galaxy.}
  \label{fig:hiprof}
\end{figure*}

\end{document}